\documentclass[natbib]{emulateapj}
\usepackage{natbib}
\usepackage{footnote}
\usepackage{amsmath}
\usepackage{hyperref}
\hypersetup
{
bookmarksopen=true,
pdftitle="The inner edge of the HZ",
pdfauthor="", 
pdfmenubar=true, 
pdfhighlight=/O, 
pdfpagemode=UseNone, 
pdfpagelayout=SinglePage, 
pdffitwindow=true, 
linkcolor=blue,
citecolor=linkcol, 
urlcolor=linkcol 
}

\shorttitle{The inner edge of the HZ}
\shortauthors{Zsom et al.}

\begin{document}

\title{Towards the Minimum Inner Edge Distance of the Habitable Zone}

\author{Andras Zsom, Sara Seager, Julien de Wit, Vlada Stamenkovi\'{c}}
\affil{Department of Earth, Atmospheric and Planetary Sciences, Massachusetts Institute of Technology, Cambridge, MA 02139, USA; \texttt{zsom@mit.edu}}

\begin{abstract}

We explore the minimum distance from a host star where an exoplanet could potentially be habitable in order not to discard close-in rocky exoplanets for follow-up observations. We find that the inner edge of the Habitable Zone for hot desert worlds can be as close as 0.38 AU around a solar-like star, if the greenhouse effect is reduced ($\sim$ 1\% relative humidity) and the surface albedo is increased. We consider a wide range of atmospheric and planetary parameters such as the mixing ratios of greenhouse gases (water vapor and CO$_2$), surface albedo, pressure and gravity. Intermediate surface pressure ($\sim$1-10 bars) is necessary to limit water loss and to simultaneously sustain an active water cycle. We additionally find that the water loss timescale is influenced by the atmospheric CO$_2$ level, because it indirectly influences the stratospheric water mixing ratio. If the CO$_2$ mixing ratio of dry planets at the inner edge is smaller than 10$^{-4}$, the water loss timescale is $\sim$1 billion years, which is considered here too short for life to evolve. We also show that the expected transmission spectra of hot desert worlds are similar to an Earth-like planet. Therefore, an instrument designed to identify biosignature gases in an Earth-like atmosphere can also identify similarly abundant gases in the atmospheres of dry planets. Our inner edge limit is closer to the host star than previous estimates. As a consequence, the occurrence rate of potentially habitable planets is larger than previously thought.

\end{abstract}


\section{Introduction}
One of the main driver of exoplanet sciences is to find habitable or inhabited exoplanets. By definition an exoplanet is habitable if liquid water, a crucial ingredient for life as we know it, is present on its surface, and if the surface temperature and pressure are such that complex organic molecules are stable \citep{Lang1986}. Habitability is inferred remotely by studying the atmosphere of exoplanets and retrieving atmospheric physical conditions such as the temperature profile and composition \citep[see e.g.,][]{Seager2000, Brown2001, Grillmair2007, Tinetti2007, Madhusudhan2009, Lee2012, Benneke2012}, or more specifically the presence of biosignature gases \citep[e.g.,][]{DesMarais2002, Segura2005, Kaltenegger2010a, Sterzik2012}. Surface properties might be retrievable if the atmosphere is optically thin in so-called window regions of the spectrum \citep[e.g.,][]{Seager2005, Fujii2011, Hu2012, Hegde2013}. 

We summarize highlights of previous work on the inner edge of the Habitable Zone going from the outermost to the innermost inner edge distance\footnote{To be habitable, a planet cannot be arbitrarily close to its host star. One reason is that any habitable planet must have, by definition, liquid water on the surface thus water vapor in the atmosphere. As water vapor absorbs the stellar light in the infrared, it increases the planet temperatures for a given incoming stellar flux, thereby, limiting the minimum distance from the star at which the planet can sustain liquid water.}. Typically an Earth-sized planet with an Earth-like atmosphere is studied, by changing one attribute (e.g., surface pressure) at a time. We describe this attribute and the level of model complexity. The pioneering work of \cite{Hart1979} placed the inner edge at 0.96 AU around a solar-like star. The recent 1D cloud-free calculations show that the inner edge is 0.92 AU assuming an Earth-like atmosphere on a super-Earth planet \citep{Kopparapu2013}. Earlier models suggest that the inner edge is at 0.88 AU for an Earth-like planet with 10 bar N$_2$ \citep{Kasting1993}\footnote{The calculation with the 10 bar N$_2$ atmosphere was not performed in the recent study.}. By maximizing the cooling effect of clouds on an Earth-like planet, the inner edge could be located as close as 0.87 AU around the Sun based on 1D models with clouds \citep{Kitzmann2010}. Using an energy balance model and varying the surface pressure between 0.3 and 3 bars, \cite{Vladilo2013} showed that the inner edge is located at 0.85 AU within the parameter range where their model is calibrated. The inner edge around a solar-like star is located at 0.77 AU for an Earth-like planet with a limited water reservoir based on 3D global circulation models \citep{Abe2011}. Global circulation models also show that tidally locked planets with a stellar insolation of $\sim$ 2400 W/m$^2$ can still be habitable due to the climate stabilizing cloud feedback \citep{Yang2013}. This insolation corresponds to an equivalent distance of 0.7 AU around a solar-like star. Finally, the inner edge could be as close as 0.46 AU around a solar-like star, if the cooling effect of clouds is maximized \citep{Selsis2007}. This means that 100\% cloud coverage is assumed with a Venus-like cloud albedo of 0.7 - 0.8, and the warming effect of clouds is neglected. However, 100\% cloud coverage and the high cloud albedo at the inner edge are not warranted (we are unaware of such model predictions), and the warming effect of clouds is not negligible \citep[see e.g.,][]{Goldblatt2011}. If these assumptions are not met, the inner edge is plausibly pushed further than 0.46 AU away from the star.

Here, we find that the inner edge is close to the host star, when non-Earth-like atmosphere conditions are considered. Super-Earth and Earth-sized planets are expected to be diverse and an extension of diversity to atmospheres must be considered in habitability studies, if we wish to probe the parameter space of habitability without an Earth-centric bias. We must first identify the key parameters influencing habitability and then establish the broadest (but plausible) range where the parameters could reside. In this work, the extreme value is adopted for some parameters, and the full range is considered for others. We use a variety of methods to estimate habitability: we calculate 1D radiative-convective temperature-pressure and mixing ratio profiles; we study the effect of atmospheric circulation on the climate using order of magnitude estimates; and we also calculate the water loss timescale\footnote{The water loss timescale defines the typical timescale over which the water reservoir of the planet is depleted due to the photodissociation of stratospheric H$_2$O molecules followed by the escape of H to space.}. All these methods are combined to identify the atmospheric scenarios that minimize the inner edge distance, and to reject scenarios where the planet is hostile to life.

The inner edge of the habitable zone is controlled mainly by the properties of the host star, the planet atmosphere's relative humidity, CO$_2$ mixing ratio, the nature of the most abundant gas, and the surface pressure. The first order effect of the host star is related to the stellar luminosity. The stellar luminosity increases with stellar mass, therefore the habitable zone is more distant around massive stars \citep{Kasting1993}. The atmospheric relative humidity is controlled by the water cycle. The CO$_2$ mixing ratio is affected by the carbon-silicate cycle as well as the water cycle (both cycles are discussed in detail in the next paragraph). The most abundant gas in the atmospheres of close-in habitable exoplanets is expected to be optically inactive around 10 $\mu m$, e.g., not a greenhouse gas \citep[including H$_2$,][]{Borysow2002}, because then it is ensured that the greenhouse warming is inefficient and the atmosphere can rapidly cool in spite of the strong stellar irradiation. The surface pressure is determined by various processes including (but not limited to) accretion from the protoplanetary disk, outgassing, atmospheric escape, accretion/erosion during impacts. Other aspects such as e.g., clouds \citep[e.g.,]{Kitzmann2010, Goldblatt2011, Zsom2012, Yang2013} or tidal heating \citep{Barnes2012} could also impact the inner edge distance. We do not model clouds because we expect low cloud fractions on hot desert worlds. But we illustrate how representative tidal heating distances compare to our derived inner edge limits.

The water cycle and the carbon-silicate cycle are the two most important processes that can alter the atmospheric composition and the surface pressure, and so we discuss them in more detail. The water cycle provides a positive climate feedback. As the surface becomes warmer, the evaporation rate of the surface waters increases and water vapor builds up in the atmosphere. As the optical depth of the atmosphere increases with the water mixing ratio, the greenhouse effect becomes stronger, promoting further warming and evaporation \cite[see e.g.,][and references therein]{Goldblatt2012}. The carbon-silicate cycle is a climate-stabilizer feedback process that operates on a $10^6$ years timescale \citep{Walker1981}. CO$_2$ is removed from the atmosphere by the interaction of weathered silicate rocks with dissolved atmospheric carbon in form of carbonate ions CO$_3$(2-) \citep{Abbot2012}: the weathered rocks are transported to oceans where abiotic or biotic processes produce carbonates XCO$_3$, (X is most commonly Ca, Mg, Fe) that sediment to the seafloor. If plate tectonics is present, those carbonates are subducted to the mantle and removed from the surface regions. CO$_2$ reforms by metamorphism in the mantle and it is outgassed by volcanism that closes the carbon-silicate cycle. If the climate cools, silicate weathering rates are reduced \citep{Walker1981}, therefore outgassed CO$_2$ builds up in the atmosphere and warms the surface to a new equilibrium temperature. The width of the Habitable Zone (HZ) is affected by these two cycles for an Earth-like planet \citep{Kasting1993, Kopparapu2013}.

We do not model the outer edge of the Habitable Zone, because even free-floating planets could be habitable \citep{Stevenson1999}. In other words, the outer edge of the Habitable Zone is currently considered at infinity. Such planets are heated from the interior (e.g., radioactive heating, or primordial heat), and must also have a thick H$_2$ atmosphere to reduce the outgoing thermal emission.

The paper's outline is the following: in Sect. \ref{sec:methods}, we first describe the initial atmosphere profiles and how the convective tropospheric temperature-pressure profile is calculated (Sect. \ref{subsec:iniprof}). Then we introduce our climate code specifically developed for this study  (Sect. \ref{subsec:clima}). We describe the influence of atmospheric parameters on the inner edge distance in Sect. \ref{subsec:ranges}. We develop atmospheric dynamics tools to assess the plausibility of best case scenarios in Sect. \ref{subsec:prec}, and estimate the loss timescale of surface waters in Sect. \ref{subsec:loss}. Our results are outlined in Sect. \ref{sec:results}, where we present our main findings and supporting arguments. In Sect. \ref{sec:discussion}, we discuss the formation and the observable signatures of hot desert worlds, and the model uncertainties. Finally we summarize our results and draw conclusions in Sect. \ref{sec:sum}.

\section{Methods}
\label{sec:methods}
Our aim is to find plausible atmospheric conditions that provide habitable surface conditions as close as possible to the host star. We do this in three steps: first, we study how various atmospheric and planetary properties influence the location of the inner edge by calculating radiative-convective equilibrium temperature-pressure profiles (Sections \ref{subsec:iniprof}, \ref{subsec:clima}, and \ref{subsec:ranges}); second, we estimate the dominant form of water precipitation (rain or snow) and the atmospheric circulation requirements for precipitation (Sect. \ref{subsec:prec}); and third, we estimate the water loss timescale to determine whether the surface water reservoir is stable on a multi-billion year timescale (Sect. \ref{subsec:loss}).

We use order of magnitude estimates and 1D atmosphere models to estimate the inner edge distance unlike \cite{Abe2011}. Although 3D global circulation models capture a wide range of atmospheric dynamical phenomena, 1D models are computationally efficient. Therefore, 1D models are suitable to study a broad range of atmospheric configurations (various surface pressures, CO$_2$ mixing ratios, relative humidities, etc.) as a function of stellar properties.

\subsection{Initial atmospheric profiles and convection}
\label{subsec:iniprof}
The T-P profile and volume mixing ratios are initialized with a convective zone that follows the moist adiabat from the surface to the tropopause, and an isothermal stratosphere. The atmosphere is described by the surface conditions (pressure, temperature, and albedo), relative humidity, CO$_2$ mixing ratio, background gas (typically N$_2$), and planetary surface gravity. Unless otherwise stated, we use 60 layers per pressure decade. The top-of-atmosphere pressure is fixed at 10$^{-4}$ bar because the atmosphere is optically thin in all spectral bands above this level. The surface pressure is varied between 0.1 and 100 bars. 

First, we establish the mixing ratios of CO$_2$ and the background gases on a dry pressure grid (without water vapor). We assume that these gases are well mixed in the atmosphere and their mixing ratios are constant as a function of pressure. This assumption is correct, if the mixing time scale in the atmosphere is much shorter than the time scale on which the sources and sinks of the atmospheric constituents operate. 

Next, the moist adiabat is built from the surface upwards given the dry surface pressure ($P_{\mathrm{surf, \, dry}}$), temperature ($T_{\mathrm{surf}}$), and relative humidity ($\Phi$) following \cite{Pierrehumbert2011}:
\begin{equation}
\frac{d \ln T}{d \ln P_{\mathrm{dry}}} = \frac{R_{bg}}{c_{p,\,bg}} \frac{1 + \frac{L}{R_{bg} T} f_{H_2O}}{1 + \left[ \frac{c_{p,\,H_2O}}{c_{p,\,bg}} + \left( \frac{L}{R_{H_2O} T} - 1 \right) \frac{L}{c_{p,\,bg} T} \right]f_{H_2O}},
\label{eq:adi}
\end{equation}
where $R_{bg}$ and $R_{H_2O}$ are the specific gas constant of the background gas and water vapor, respectively; $c_{p,\,bg}$ and $c_{p,\,H_2O}$ are the specific heat capacity of the background gas and water vapor, respectively; $L$ is the latent heat of water vapor, and $f_{H_2O}$ is the water vapor mass mixing ratio:
\begin{equation}
f_{H_2O} = \frac{m_{H_2O}}{m_{bg}} \frac{ \Phi P_{\mathrm{sat}}(T)}{ P_{\mathrm{dry}}},
\end{equation}
where $m_{bg}$ and $m_{H_2O}$ are the molecular masses of the background gas and water vapor, respectively, and $P_{\mathrm{sat}}(T)$ is the saturation vapor pressure at temperature $T$ calculated according to the Smithsonian Meteorological Tables. The moist adiabat is followed until the temperature drops below the prescribed initial stratosphere temperature. The stratospheric water vapor mixing ratio is constant and it is determined by the tropopause temperature and relative humidity. Finally, the total atmospheric pressure is calculated as the sum of the water vapor partial pressure and the dry pressure. 

\subsection{Radiative-convective equilibrium profiles}
\label{subsec:clima}
We develop a specific numerical approach to calculate the T-P profiles: we fix the surface temperature and determine the semi-major axis where the atmosphere is in radiative equilibrium. Our approach is untypical in a sense that commonly the exoplanet's semi-major axis is fixed and the corresponding surface temperature is calculated \citep[see e.g., the EXO-P model of][]{Kaltenegger2010b}. But, as we want to determine the equilibrium planet-star separation for given surface temperature and pressure, our method is better suited for this study.

The equilibrium T-P profile is determined by iterating the following steps: 
\begin{enumerate}
\item We calculate the fluxes propagating through each atmospheric layer in several wavelength bands; we treat the thermal emission of the exoplanet's surface/atmosphere and the stellar irradiation separately, and evaluate the stellar irradiation fluxes at an arbitrary distance.
\item We determine the semi-major axis where the top-of-atmosphere fluxes are in equilibrium using the Bond albedo of the exoplanet and the outgoing long wave radiation.
\item The stellar irradiation fluxes are rescaled such that the incoming stellar flux corresponds to the value received at the new semi-major axis.
\item The stratosphere temperature profile is updated based on the net fluxes propagating through the layers.
\item The tropopause height is updated if necessary. 
\end{enumerate}
The iteration stops, when the T-P profile reaches equilibrium. These steps are described in the following.

\subsubsection{Radiative transfer}
Fluxes are calculated in a limited number of wavelength bands for computational efficiency. The bands range between 0.2 to 200 microns, and they are linearly spaced in $\log(\lambda)$ (i.e., the spectral resolution is constant). Such spacing ensures that Planck functions of various temperatures are resolved with the same power. We use 400 wavelength bands. The sufficient number of bands were determined by convergence tests where the outgoing IR flux of a test atmosphere was monitored as the number of bands is gradually increased.

We calculate the optical depth of each layer in each wavelength band given the mixing ratio profiles. Various absorption and scattering sources could contribute to the optical depth of a layer. These are the greenhouse gas absorption (discussed in the next paragraph), Rayleigh scattering \citep{Harvey1998, Weber2003, Sneep2005}, collision induced absorption (CIA) of N$_2$ \citep{Borysow2002} and of CO$_2$ \citep{Gruszka1997}, and continuum absorption of water and CO$_2$ \citep[following the description of][Sect. 4.4.8]{Pierrehumbert2011}. 

We use the line-by-line HITRAN2012 database \citep{Rothman2009}\footnote{The article describing the 2012 edition of HITRAN will appear in a Special Issue of the Journal of Quantitative Spectroscopy and Radiative Transfer that is expected to come out later this year.} and the exponential sums formalism \citep{Wiscombe1977, Pierrehumbert2011} to determine the band-averaged transmission through an atmospheric layer due to greenhouse gas absorption. The high resolution line-by-line absorption coefficient is calculated at a pressure and temperature grid covering typical values encountered in our atmosphere profiles. We use the semi-analytical Voigt line profile of \cite{Liu2001}, and determine the Voigt line width using the Doppler and Lorentz line widths \citep{Olivero1977}. The histogram of absorption coefficients are pre-calculated and tabulated, and these tables are read and interpolated at run-time to the pressure and temperature values of each layer. 

Once the optical depth of the atmosphere is known, fluxes propagating through the layer interfaces are calculated. We use the two-stream approximation and adopt the quadrature method for the stellar illumination, and the hemispheric mean approximation for the thermal radiation \citep{Toon1989}. The thermal flux of the planetary surface and atmosphere are treated separately from stellar irradiation. The distinction of the two fluxes is important to determine the equilibrium distance of hot inner edge planets orbiting low temperature M stars. The outgoing thermal emission of the planet and atmosphere might overlap with the reflected stellar light in the near-IR. However, it is necessary to determine the relative contribution of these fluxes to calculate the equilibrium distance of the planet (see Eq. \ref{eq:dist}).

\subsubsection{Climate model}
The goal of the climate code is to iterate the temperature and water vapor profiles to a radiative-convective equilibrium based on the fluxes propagating through the layer interfaces. This is achieved in three stages. First, we determine the planet-star separation where the top-of-atmosphere incoming and outgoing fluxes are in equilibrium, and rescale the stellar fluxes using the new distance. Then we update the stratospheric temperature profile, and finally the tropopause height is determined. 

The distance where the top-of-atmosphere fluxes are balanced is calculated as
\begin{equation}
OTE = \frac{1}{4} (1 - \alpha) F_{in}(a),
\label{eq:dist}
\end{equation}
where $OTE$ is the top-of-atmosphere outgoing thermal emission, $\alpha$ is the Bond albedo of the planet, and $F_{in}$ is the top-of-atmosphere incoming stellar irradiation. As the T-P and mixing ratio profiles are assumed constant within one iteration, the outgoing thermal emission is independent of distance $a$. The albedo (the ratio of top-of-atmosphere outgoing and incoming stellar fluxes) is also distance-invariant, and it is evaluated at an arbitrary distance, typically where the incoming stellar flux equals the Solar constant. Therefore, the unknown in Eq. \ref{eq:dist} is $F_{in}$ and the distance $a$ where $F_{in}$ is received at the top of the atmosphere. Once the equilibrium incoming stellar flux is determined, the downwelling and upwelling stellar fluxes are scaled. 

We update the stratospheric temperature in the next step. We adopt a numerically efficient and simple approach to update the T-P profile: the temperature of a layer is increased by $dT$, if the net flux propagating through the layer ($dF_i$ - the difference between incoming and outgoing fluxes) is positive, and the temperature decreases otherwise. The temperature step ($dT$) is 0.5 K in our simulations. Our method differs from traditional approaches that calculate the derivative $dT_i / dt$ (change of temperature over time) in each atmospheric layer $i$:
\begin{equation}
\frac{dT_i}{dt} = \frac{g}{c_{pi}} \frac{dF_i}{dP_i},
\label{eq:dTdt}
\end{equation}
where $g$ is the gravitational constant, $c_{pi}$ is the total heat capacity of the layer, $dF_i$ is the net flux propagating through the layer, $dP_i$ is the pressure difference at the top and bottom of the layer. If the net flux is negative, the layer cools. The time step ($dt$) is adaptively determined by setting an upper limit on $dT_i$ ($dT_{max}$). We find that although the traditional approach properly captures how the temperature profile evolves in time, it converges slowly if the tropopause pressure level is orders of magnitude larger than the top-of-atmosphere pressure level. Furthermore, it is unimportant how the temperature profile evolves in time from an arbitrary initial state, as the goal is to determine an equilibrium profile.

Finally, the tropopause height and the stratospheric water mixing ratio are updated in the climate code. The lapse rate between the tropopause layer and the layer above is calculated. The tropopause height is increased by one layer if the lapse rate is steeper than the moist adiabat. The tropopause height is decreased by one layer if the lapse rate is larger than 0 K/km (i.e., the layer above the tropopause is warmer). As the tropopause temperature changes, so do the tropopause and stratospheric water mixing ratios. 
\\\\
The steps outlined in the previous two sections, namely radiative transfer, determining the equilibrium distance, stratosphere temperature, tropopause, and water mixing ratio updates are iteratively performed until the temperature profile relaxes to equilibrium. Our convergence criteria measures the temperature change of each layer for $n$ iterations. Convergence is reached, if the temperature change is less than $2\, dT$ in all of the layers during $n$ successive iterations, where $n$ is typically 10. Model validation and tests are described in App. \ref{app:modval}.

The temperature-pressure profiles described in Sect. \ref{sec:results} have some general characteristics: a convective zone close to the surface, a cold tropopause, and inversion in the stratosphere. We assume that regions close to the surface are convective. The stratospheric inversion occurs because water vapor absorbs the incoming stellar radiation in the near infrared. CO$_2$ does not have absorption bands at such short wavelengths, but it emits in the infra red, thus CO$_2$ acts as a stratospheric coolant \citep{Clough1995}.

\subsection{The ranges of model parameters, and their effect on the inner edge}
\label{subsec:ranges}
We provide a qualitative description on how the various atmospheric parameters influence the outgoing thermal flux, the albedo of the planet, and thus the inner edge distance in this section. In general, increasing the outgoing thermal emission and/or the albedo of the planet moves the inner edge closer to the star because more incoming stellar flux is necessary to achieve radiative-convective equilibrium. The most important effects of each parameter is described in detail in Sect. \ref{sec:results}, while other parameters are fixed during the simulations. 

\paragraph{Stellar type}
We perform simulations using various stellar types ranging from M6V to G0V (or stellar masses ranging from 0.09 and 1.1 M$_\odot$, respectively). We use the ``BT\_Settl'' model spectra grid of \cite{Allard2003, Allard2012} \citep[to make comparison easier with][]{Kopparapu2013}. The stellar effective temperatures range between 2700 K and 6000 K with a step of 300 K. We choose model spectra with solar metallicity \citep{Asplund2009} and assume a $\log g$ of 4.5. The stellar mass and luminosity is assigned to the stellar spectra from the stellar evolution models of \cite{Baraffe1998, Baraffe2002} assuming an age of $5\times10^9$ yrs.

The spectral energy distribution of the star has several effects on planetary climate and habitability. The first order effect of stellar type is related to the luminosity of main sequence stars. An exoplanet has to be much closer to an M dwarf than to an F star to receive the same total flux at the top of its atmosphere. However, the spectral energy distribution of stars has higher order effects on the inner edge distance. These effects become apparent, if the top-of-atmosphere fluxes are compared at the inner edge around various stars \citep{Kopparapu2013}. As an example, the spectral energy distributions of low mass stars peak at longer wavelengths where Rayleigh scattering is less efficient, thus the planetary albedo is reduced. Furthermore, a larger fraction of the incoming stellar radiation is directly absorbed by the atmosphere around low mass stars because the greenhouse gases such as H$_2$O and CO$_2$ absorb in the near IR. To a smaller extent, the UV flux of the host star (especially the extreme UV) influences the composition of the stratosphere, which in turn can slightly influence the surface climate \citep[see Fig. 4a of][]{Rugheimer2013}.

\paragraph{Surface gravity}
The effects of surface gravity are two-fold. If the surface pressure is fixed, the cumulative optical depth of the atmosphere is somewhat reduced at all wavelengths for a larger surface gravity because the atmospheric scale height and thus the column density of greenhouse gases are reduced. On one hand, this increases the outgoing thermal emission of the exoplanet and shifts the inner edge closer to the star. On the other hand, more stellar flux reaches the surface because less stellar flux is absorbed by the atmosphere, and the column density of Rayleigh scattering background gas is reduced. If the surface albedo is low, the Bond albedo of the planet is reduced and the inner edge is pushed further away from the star. If the surface is reflective, the inner edge is pushed closer to the star. Therefore the albedo-effect of surface gravity depends on the relative contributions of two effects to the Bond albedo: Rayleigh scattering, and surface reflection. The mass and radius of an exoplanet also influences how much atmosphere the planet can retain \citep[see e.g.,][]{Lammer2009}, and what the volatile content of the mantle is that can potentially be outgassed. 

We use values between $g = 5$ m/s$^2$ for a planet somewhat larger than Mars, 10 m/s$^2$ for an Earth-like planet, and 25 m/s$^2$, which is a typical surface gravity of a 10 M$_\oplus$ super-Earth. 

\paragraph{Surface pressure}
The effect of surface pressure on the inner edge distance is two-fold. Larger surface pressure increases the albedo of the planet due to Rayleigh scattering. However, an increasing surface pressure also raises the column density of the greenhouse gases and thus the cumulative optical depth of the atmosphere for a fixed mixing ratio. Therefore, the effect of surface pressure on the inner edge distance cannot be readily estimated, thus we treat it as a free parameter. We consider planets with 0.1, 1, 10, and 100 bars of surface pressure. 

\paragraph{CO$_2$ mixing ratio}
The lower the CO$_2$ mixing ratio is, the closer the inner edge of the HZ is to the star. CO$_2$ in combination with water vapor is a very effective greenhouse gas because these gases have non-overlapping absorption bands in the infrared. If CO$_2$ was removed from the atmosphere of Earth, the global average temperature would be below the freezing point of water \citep{Pierrehumbert2011}. Therefore it is crucial to calculate how the mixing ratio of CO$_2$ influences the inner edge distance.

The CO$_2$ mixing ratio is a free parameter in our study owing to the uncertainties in the outgassing and deposition rates. The mixing ratio of CO$_2$ is varied between $10^{-5}$ and $10^{-1}$. We do not consider higher CO$_2$ mixing ratios because such planets are expected to be unhabitable close to the host star due to the greenhouse effect of CO$_2$. Thus, an assumption in our model is that some process (e.g., the carbon-silicate cycle) keeps the atmospheric CO$_2$ at bay. If this assumption is not met, the atmosphere becomes CO$_2$-dominated like the atmosphere of Venus.

\paragraph{Relative humidity}
The relative humidity is a key parameter in our study because it has a strong influence on the inner edge distance. Unfortunately it is exceedingly difficult to model the relative humidity profile of an exoplanet because it is determined by complex 3D processes such as atmospheric circulation, cloud formation, and precipitation. The water cycle is also influenced by the spatial distribution of liquid water on the planetary surface. We do not attempt to model these effects. Therefore, the relative humidity is a free parameter in our study and it is changed between 0.01\% and 70\%. The nominal value is 1\% which we determine by estimating the dominant form of precipitation (see Sect. \ref{subsec:prec}).

\paragraph{Surface temperature}
The inner edge of dry planets is set by the boiling point of water because dry planets do not enter the moist runaway greenhouse stage (even at these high surface temperatures, see Sect. \ref{sec:results}). The boiling point, thus the surface temperature at the inner edge is determined by the surface pressure and the relative humidity:
\begin{equation}
T_{\mathrm{surf}} = \max(T:P_{\mathrm{surf}} + \Phi P_{\mathrm{sat}}(T) > P_{\mathrm{sat}}(T)\,),
\label{eq:Tsurf}
\end{equation}
where $P_{\mathrm{surf}}$ is the dry surface pressure, $\Phi P_{\mathrm{sat}}(T)$ is the partial pressure of water on the surface. The inner edge surface temperature of Earth could be 410 K for a dry surface pressure of 1 bar, and the surface relative humidity is 70$\%$. If Earth had 1$\%$ relative humidity and all other properties kept constant, the inner edge surface temperature would be lower, 373 K because the partial pressure of water is reduced. 

The maximum surface temperature is also limited by the chemical stability of complex organic molecules. If only Eq. \ref{eq:Tsurf} is used to calculate the inner edge surface temperature, its allowed range is between 273 K and 647 K given by the triple and critical points of water, respectively. Can life exist at temperatures as high as 647 K? On Earth, organisms grow at temperatures up to 395 K \citep{Kashefi2003, Takai2008}. However, the DNA and amino acids become chemically unstable generally at temperatures above 500 K \citep{Lang1986}. Therefore we limit the surface temperature to a maximum of 500 K. We note that this is a globally averaged surface temperature because we use a 1D climate model. The surface temperature on the globe is expected to be scattered around the mean value as shown by global circulation models \citep[see e.g.,][]{Leconte2013}. Therefore the surface temperature must remain well below 500 K at a substantial part of the planet (see Sect. \ref{subsec:prec}). 

These inner edge surface temperatures differ from previous estimates. Most previous studies assumed that the troposphere is saturated with water vapor, the tropopause and stratosphere temperatures are set to 200 K, and Earth's water reservoir is present on the surface \citep{Kasting1993, Kopparapu2013}. Under these conditions, the onset of the moist greenhouse stage is one way to define the inner edge of the habitable zone on an Earth-like planet that occurs at surface temperatures of 340 K (below the boiling point of water). The moist greenhouse stage starts when the water loss timescale becomes comparable to the age of Earth. We show in Sect. \ref{sec:results} that the water loss timescale is typically long on dry planets even when the surface temperature is close to the boiling point, because the stratosphere is dry.

\paragraph{Surface albedo}
We assume that our exoplanets are hot and dry, thus most of the surface at low zenith angles is possibly covered by deserts or barren rocky surfaces. The average surface albedo of Earth is 0.15 \citep[Chapter 5 of][based on ERBE measurements]{Pierrehumbert2011}, and the main contributors are water, land, vegetation, and ice/snow surfaces. As a reference, the albedo of liquid water is 0.05 \citep{Clark2007}, and the albedo of the Sahara is 0.4 \citep{Tetzlaff1983}. We consider surface albedos between 0 and 0.8. The high value is hypothetical (maybe unrealistic) upper limit because typical rock-forming minerals have a low albedo \citep{Clark2007}. A realistic surface albedo might be 0.2 and we choose this as our nominal value (see Table \ref{table:para}). We find that although higher surface albedos of low surface pressure atmospheres have an impact on the inner edge distance, the surface reflection becomes less important at high surface pressures because most of the incoming stellar flux is scattered back to space by Rayleigh scattering, or absorbed directly by the atmosphere. At 100 bars, typically less than 1$\%$ of the stellar flux reaches the surface. If the surface pressure is low, the surface albedo influences the tropopause height because the convective zone is driven by the absorption of stellar light at the surface.

\paragraph{Nominal parameter values} The parameter values of our nominal simulation is shown in Table \ref{table:para}. The most important effect of these parameters is shown in Sect. \ref{sec:results}. We typically change one parameter at a time, while the other parameters are held fixed at the value given in Table \ref{table:para}. In comparison, \cite{Abe2011} assumes a surface gravity of 10 m/s$^2$, a surface pressure of 1 bar, an Earth-like N$_2$/O$_2$ dominated atmosphere with a present Earth CO$_2$ mixing ratio, a minimum of 50\% relative humidity, and a surface albedo of 0.3.

\begin{table}
\caption{Overview of parameter values in the nominal simulation. These values correspond to a 10 Earth mass super Earth with a nitrogen-dominated atmosphere and low relative humidity.}
\label{table:para}      
\begin{center}   
\small
\begin{tabular}{l c}        
\hline\hline
Parameter & Value \\ 
\hline
surface gravity & $g_{\mathrm{surf}}$ = 25 m/s$^2$ \\
surface pressure & $P_{\mathrm{surf}}$ = 1 bar \\
CO$_2$ mixing ratio & $X_{CO_2}$ = 10$^{-4}$ \\
relative humidity & $\Phi$ = 1\% \\
surface temperature & $T_{\mathrm{surf}}$ = 370 K \footnote{The surface temperature is set by the boiling point of water and it is a function of relative humidity and surface pressure (see Eq. \ref{eq:Tsurf}).}\\
surface albedo & $a_{\mathrm{surf}}$ = 0.2 \\
background gas & N$_2$ \\
\hline
\end{tabular}
\end{center}
\end{table}

\subsection{The interplay of atmospheric circulation, and the hydrological cycle}
\label{subsec:prec}
It has been hypothesized that dry planets could trap their water reservoir permanently on cold areas like the poles and the night side \citep{Menou2013,Leconte2013}. Although such a scenario can provide habitable regions on the planet (ice sheets flowing towards the day side and melting there), the habitable area is rather limited. It needs to be studied whether future observatories \citep[such as the \textit{James Webb Space Telescope},][]{Clampin2010} will be able to pick up biosignature gases from planets that are only partially habitable.

One way to enlarge the habitable region is to consider scenarios where precipitation occurs predominantly in the form of rain. The water reservoir is still limited to the night side or poles. However, most of the surface waters are in liquid form, thus a larger fraction of the planet's surface is habitable, and the water cycle is not broken.

We estimate that if the atmospheric relative humidity is 1\%, rain is the dominant form of precipitation for a variety of surface pressures and temperatures. We use the dew point temperature to estimate the dominant form of precipitation. The dew point temperature ($T_d$) is given by the globally averaged temperature of the air above the surface and the relative humidity. As a point of reference, if the globally averaged surface temperature is 370 K, and the relative humidity is 1$\%$, the parcel has to cool down to $T_d = 280$ K to reach saturation. The dew point temperature for the same relative humidity value is 337 K, if the surface temperature is 500 K. Generally, dew point temperatures above 273 K ensure that liquid water precipitation is present on the planet. If the dew point temperature is below 273 K, the dominant form of precipitation could be snow.

Cloud formation by vertical mixing is not considered because precipitation from cumulus clouds might not reach the surface (called virga). If condensation occurred e.g., on the upwelling part of the Hadley circulation around the equator, it is uncertain whether rain drops reach the surface. It is likely that the drops evaporate as they fall through gradually warmer and drier regions of the troposphere. 

Large temperature differences are necessary to initiate condensation and precipitation. Therefore, the goal of this section is to study the properties of atmospheric circulation that result in a high temperature difference, and see whether the circulation pattern is plausible. We distinguish two limiting cases: slowly-rotating planets, and fast rotators like Earth. The distinction is necessary because different circulation regimes dominate the heat transfer depending on the rotation period.

\subsubsection{Slow rotators}
\label{sec:slowrot}
The equator-to-pole temperature difference is typically small on planets with long periods (several tens to hundreds of days) because the Hadley cell extends all the way to the poles. As Hadley circulation transports heat very efficiently, the equator-to-pole temperature difference is expected to be small \citep{Held1980, Seager2011}. However, the day-night temperature difference can be significant, if the atmosphere efficiently cools on the night side. In the most extreme case, the atmosphere could collapse if the temperature falls below the condensation point of nitrogen on the night side \citep[see e.g.,][]{Joshi1997, Heng2012}

The speed of the zonal surface winds and the radiative cooling timescale determine the day-night temperature difference. Using the dew point temperature ($T_d$), and the radiative cooling timescale of the atmosphere, the zonal advection timescale is estimated \citep[][Eq. 11]{Showman2002}:
\begin{equation}
\tau_{\mathrm{zonal}} =  - \tau_{\mathrm{rad}} \log \left( 1 - \frac{T_\mathrm{surf} - T_d}{T_\mathrm{surf}} \right),
\end{equation}
where $\tau_{\mathrm{rad}}$ is the radiative timescale of the atmosphere:
\begin{equation}
\tau_{\mathrm{rad}} = \frac{\Delta P}{g_{\mathrm{surf}}} \frac{c_p}{4 \sigma T_\mathrm{surf}^3},
\label{eq:tau_rad}
\end{equation}
where $\Delta P$ is the pressure difference at the top and bottom of the layer above the surface, $c_p$ is the heat capacity of the layer, and $\sigma$ is the Boltzmann constant. The zonal wind speed necessary to reach $T_d$ is 
\begin{equation}
u_{\mathrm{zonal}} = R_p / \tau_{\mathrm{zonal}},
\label{eq:zonalwind}
\end{equation}
where $R_p$ is the planet radius. 

We accept exoplanet scenarios for which the estimated zonal wind necessary to preserve the required temperature difference is large. The reason is that large temperature differences drive strong winds, but strong winds reduce temperature differences. If the estimated $u_{\mathrm{zonal}}$ is large, it is expected that the actual zonal wind speed is smaller than $u_{\mathrm{zonal}}$ because the friction between the surface and the atmosphere slows the winds. Therefore temperature gradients sufficiently large to initiate precipitation can be sustained over the globe. We use the mean surface wind speed of Earth as a reference value \citep[7 m/s,][]{Capps2008} above which we consider the circulation properties to be plausible.

\subsubsection{Fast rotators}
\label{sec:fastrot}
The day-night temperature difference on fast rotators is small, but the equator-to-pole temperature difference can be significant. The Hadley cell cannot extend all the way to the poles on fast rotators, such as Earth \citep{Held1980}. Instead, heat is transported from the maximum latitude of the Hadley cell to the poles by baroclinic instability \citep{Stone1978}. In energy balance models, baroclinic instability is approximated as a large scale diffusion process \citep[see e.g.,][]{Held1999,Spiegel2008}. We also take advantage of this formalism and calculate the diffusion coefficient necessary to create a typical temperature difference of $T_\mathrm{surf} - T_d$. The actual equator to pole temperature difference is possibly larger, as $T_\mathrm{surf}$ is a global average temperature, not the temperature at the substellar point.

The diffusion equation of the energy balance model is \citep{Seager2011}:
\begin{equation}
c_p \frac{\partial T(\phi)}{\partial t} = \nabla (c_p D \Delta T) + S(\alpha,\phi) - OTE,
\label{eq:D}
\end{equation}
where $T$, $D$, $S$, $\phi$, and $\alpha$ are the temperature, diffusion coefficient, absorbed stellar flux, latitude, and the albedo, respectively. An analytical solution exists for Eq. \ref{eq:D}, if the albedo, $D$, $c_p$ are constants, the $OTE$ is fitted as $A + BT$, and if the stellar flux is parameterized as a constant plus a term proportional to the Legendre polynomial $P_2(\cos(\phi))$ \citep{Held1999}. The diffusion coefficient expressed from the analytical solution is:
\begin{equation}
D = \frac{\left( \frac{T_\mathrm{surf}}{T_\mathrm{surf} - T_d}-1\right)B R_p^2}{6 c_p}.
\label{eq:diff}
\end{equation}
The parameter $B$ is determined by calculating radiative-convective equilibrium profiles with surface temperatures 1 K above and below the nominal $T_{\mathrm{surf}}$, and fitting the $OTE$. Our reference value for $D$ is the diffusion coefficient that reproduces the zonally averaged equator-to-pole temperature profile of Earth \citep[$10^6$ m$^2$/s,][]{Suarez1979}. Following a similar argument as in the previous section, if an atmospheric scenario requires $D < 10^6$ m$^2$/s to preserve the necessary temperature difference, we reject the scenario.

\subsection{The loss timescale of surface waters}
\label{subsec:loss}
The water reservoir of the planet is gradually depleted because the stellar UV radiation dissociates the stratospheric water vapor, and hydrogen subsequently escapes to space. The goal of this section is to estimate the loss timescale of the exoplanet's liquid reservoir assuming diffusion-limited escape, and we follow the description of \cite{Kasting1993}. If the water loss time is too short, the planet could loose its water reservoir before life develops. On the other hand, if the timescale is on the order of several billion years, it is safe to assume that the surface waters are stable and long-lasting.

The most important reason to calculate radiative-convective T-P profiles is to reliably estimate the stratospheric water mixing ratio and through that the diffusion-limited escape of H$_2$O. The tropopause pressure, temperature and relative humidity set the stratospheric water mixing ratio. It is expected that both the tropopause pressure and temperature varies with atmospheric and stellar parameters. Often it is assumed that the stratosphere has a constant 200 K temperature \citep{Kasting1993, Kopparapu2013}. However, the water loss timescale might be inaccurate as the tropopause properties are not self-consistently calculated. We note that a constant 200 K stratosphere temperature does not significantly affect the $OTE$ and thus the radiative equilibrium distance of the planet.

The water loss timescale is calculated in the following way. We estimate that our hot desert worlds have a 100 times smaller liquid water reservoir than Earth \citep{Shiklomanov1993}, thus the total mass of liquid water reservoir is $1.4\times 10^{22}$ g, which amounts to $N_{\mathrm{H_2O}} = 4.6\times 10^{44}$ H$_2$O molecules. The water loss timescale is calculated as 
\begin{equation}
\tau_{w} = \frac{N_{\mathrm{H_2O}}}{A_\mathrm{surf} n^{\mathrm{top}}_{\mathrm{H_2O}} V_{\mathrm{diff}}},
\label{Eq:loss}
\end{equation}
where $A_\mathrm{surf}$ is the surface area of the exoplanet, $n^{\mathrm{top}}_{\mathrm{H_2O}}$ is the number density of H$_2$O molecules at the top of the atmosphere, and $V_{\mathrm{diff}}$ is the diffusion velocity calculated according to \cite{Hu2012a} (see their Eqs. 3 and 20). If the top-of-atmosphere temperature is 200 K, and the water vapor mixing ratio is $3\times 10^{-5}$ on a planet with 10 m/s$^2$ surface gravity and one Earth radius, the water loss timescale is $5.2\times 10^9$ years, in close agreement with the calculations of \citet{Kasting1993, Kopparapu2013}.

\section{Results}
\label{sec:results}
We demonstrate that exoplanets can be habitable much closer to their host star than previously estimated. For example, an exoplanet can be habitable as close as 0.38 - 0.59 AU from a solar-like star given favorable atmospheric properties and surface albedos of 0.8 and 0.2, respectively. Hot and dry desert worlds with 1\% relative humidity, a broad range of CO$_2$ mixing ratios and surface pressures have sufficiently long water loss timescales, such that the liquid water reservoir would not be lost over a short period of time. The atmospheric circulation estimates show that liquid precipitation can plausibly operate under a variety of atmosphere scenarios, thus a significant fraction of the planet's surface could be habitable.

\subsection{Relative humidity as the main controlling factor}
\label{subsec:relhum}
\begin{figure}
\centering
\includegraphics[width=0.49\textwidth]{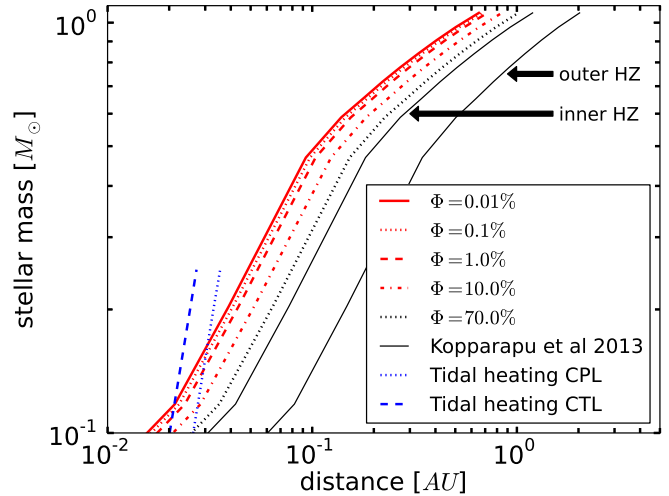}
\caption{The inner edge of the Habitable Zone for various relative humidity values. Other atmospheric parameters are described in Table \ref{table:para}. The x axis is the planet-star separation, the y axis shows the mass of the host star. The inner edge distance is small for low relative humidity. The inner and outer edge Habitable Zone limits of \cite{Kopparapu2013} are indicated with black solid lines as a comparison. Exoplanets on eccentric orbits ($e = 0.2$) located left from the blue lines could experience significant tidal heating and become unhabitable as a result. The blue dotted and dashed curves illustrate the critical tidal heating semi-major axes for dry worlds using the constant phase lag and constant time lag models of \cite{Barnes2012}, respectively. \label{fig:relhum_HZlim}}
\end{figure}

The inner edge of the Habitable Zone for dry planets with a relative humidity of 1\% can be as close as 0.59 AU around a solar-like star. The relative humidity of the atmosphere is one of the most important factors controlling how close a planet can be to a star and still maintain surface liquid water. We consider tropospheric relative humidity values between 0.01\% and 70\%. If the relative humidity is 1\% or larger, the dominant form of precipitation is expected to be rain (as shown in Sec. \ref{subsec:prec}). If the relative humidity is lower than 1\%, the necessary temperature difference (from day to night side or from equator to pole) to initiate condensation would be too high and precipitation would occur in the form of snow. The inner edge of the habitable zone for planets with various relative humidity levels is shown in Fig. \ref{fig:relhum_HZlim}. In the following we discuss the top of atmosphere radiative fluxes: the outgoing thermal emission (OTE) and the albedo of the planet.

\begin{figure*}
\centering
\includegraphics[width=0.49\textwidth]{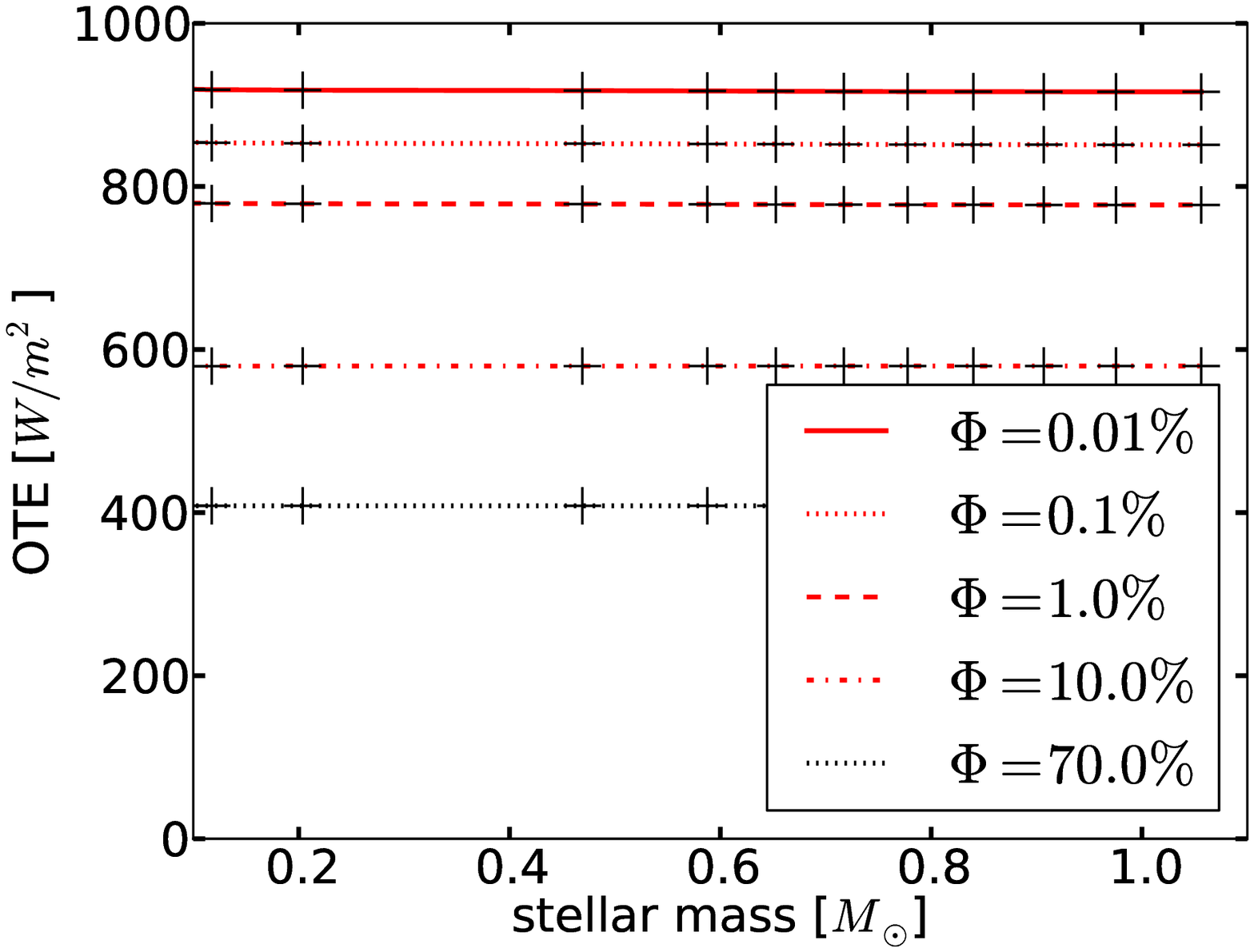}
\includegraphics[width=0.49\textwidth]{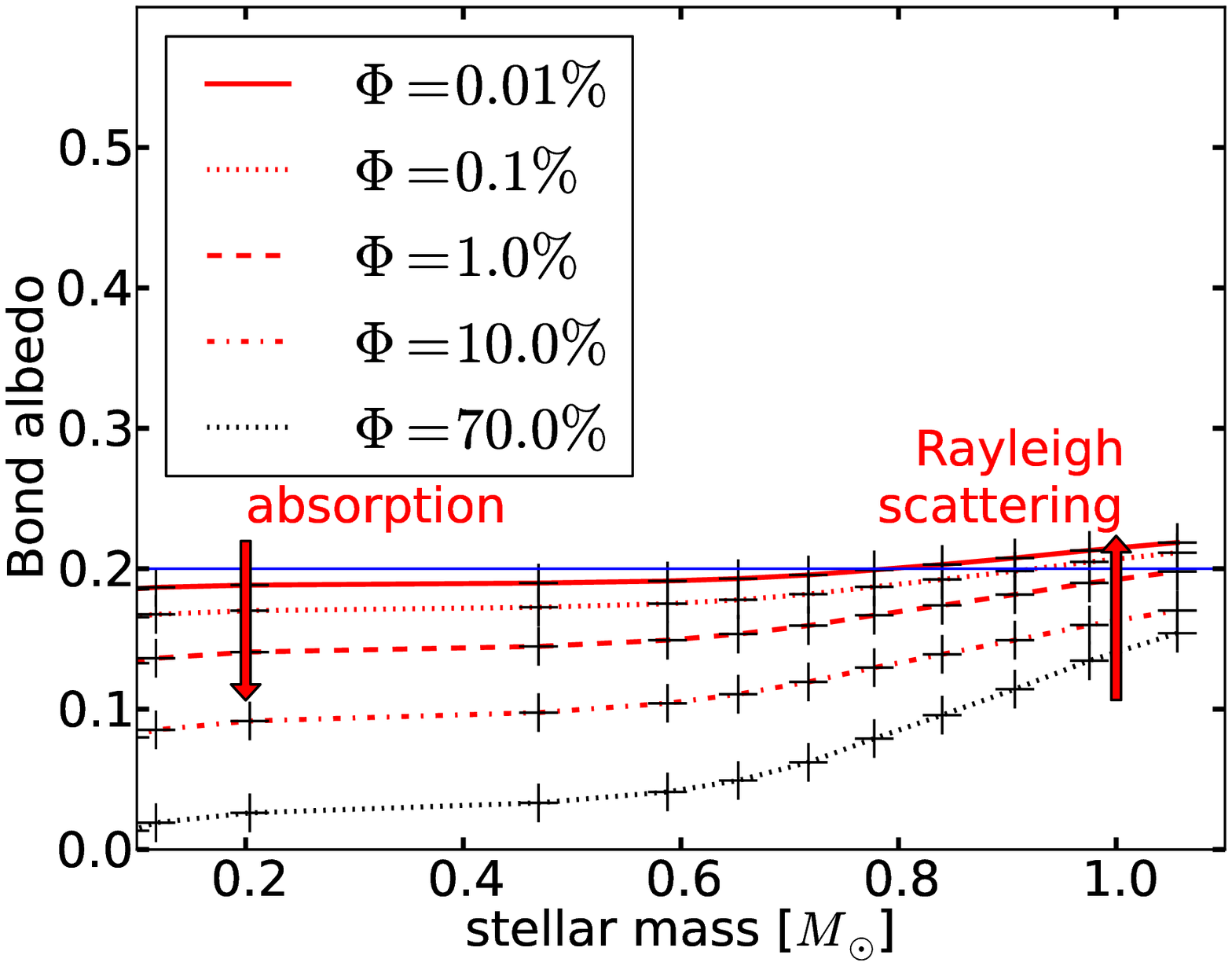}
\caption{\emph{Left:} The outgoing thermal flux for various relative humidity values. The x axis shows the mass of the host star. \emph{Right:} The Bond albedo of exoplanets for various relative humidities. The blue solid line indicates the surface albedo of 0.2 as a reference. Other atmospheric parameters are described in Table \ref{table:para}. \label{fig:relhum_OTE_alb}}
\end{figure*}

\begin{figure}
\centering
\includegraphics[width=0.49\textwidth]{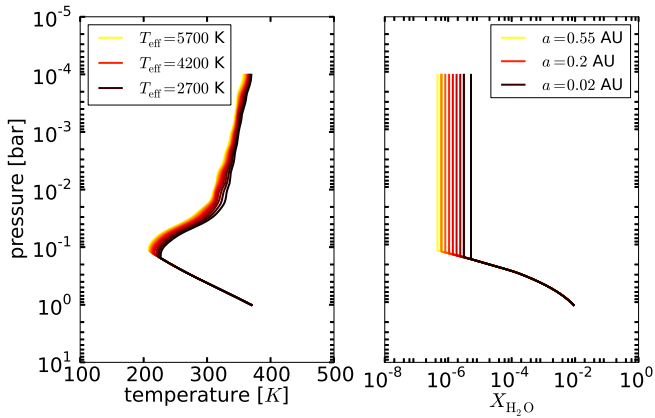}
\caption{The temperature and water mixing ratio profiles of the nominal atmosphere with parameters given in Table \ref{table:para}. The color coding represents the T-P and water mixing ratio profiles of exoplanets orbiting various host stars. The effective temperatures of three selected host stars are indicated in the legend on the left side. The legend on the right side shows the inner edge distance of the exoplanet around the host stars. The color coding is identical on all figures showing T-P and water mixing ratio profiles. \label{fig:relhum_prof_0.01}}
\end{figure}

The inner edge distance is close to the host star mainly because the outgoing thermal emission is high if the atmosphere is dry. In other words, the emitted thermal radiation escapes easily, if the greenhouse gas concentrations are low. The $OTE$ does not depend on the stellar type (see Fig. \ref{fig:relhum_OTE_alb}a); it is mainly influenced by the pressure-temperature profile and the greenhouse gas concentrations (shown in Fig. \ref{fig:relhum_prof_0.01} for the nominal atmosphere of Table \ref{table:para}).Ê

The second factor influencing the inner edge distance is the albedo of the planet. The planetary albedo is strongly influenced by the host star's spectral energy distribution, and the relative humidity \citep{Marley1999}. Low mass stars radiate at long wavelengths where absorption by greenhouse gases such as water and CO$_2$ is present. Therefore, the incoming stellar light is taken up by the atmosphere before it reaches the surface. The larger the relative humidity, the more stellar light is absorbed (see Fig. \ref{fig:relhum_OTE_alb}b). On the other hand, massive stars radiate at short wavelengths where Rayleigh scattering dominates. Therefore the planetary albedo around massive stars can be larger than the surface albedo.

\subsection{CO$_2$ mixing ratio influencing the water loss timescale}
The atmospheric CO$_2$ mixing ratio influences the tropopause temperature, which itself affects the water loss timescale. We discuss the effects of CO$_2$ on habitability from two perspectives: in this work, changes in the CO$_2$ level slightly modifies the planet's radiative equilibrium distance (aka. the inner edge distance) because the surface temperature is fixed. The other perspective is to fix the planet's semi-major axis and investigate the influence of CO$_2$ on the climate. In the latter case, the CO$_2$ mixing ratio influences the surface temperature. We shortly discuss the second perspective at the end of this section.

\begin{figure*}
\centering
\includegraphics[width=0.49\textwidth]{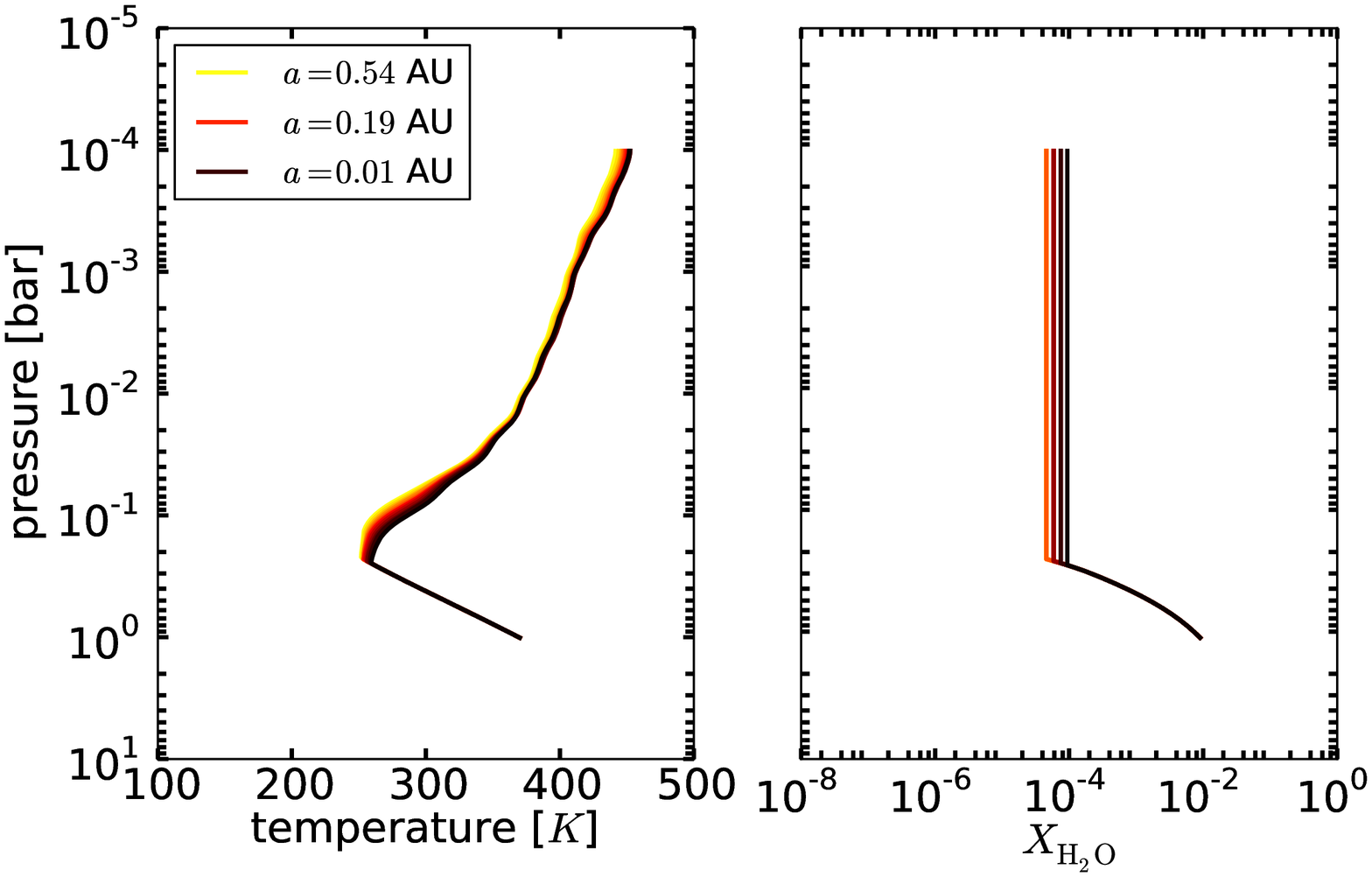}
\includegraphics[width=0.49\textwidth]{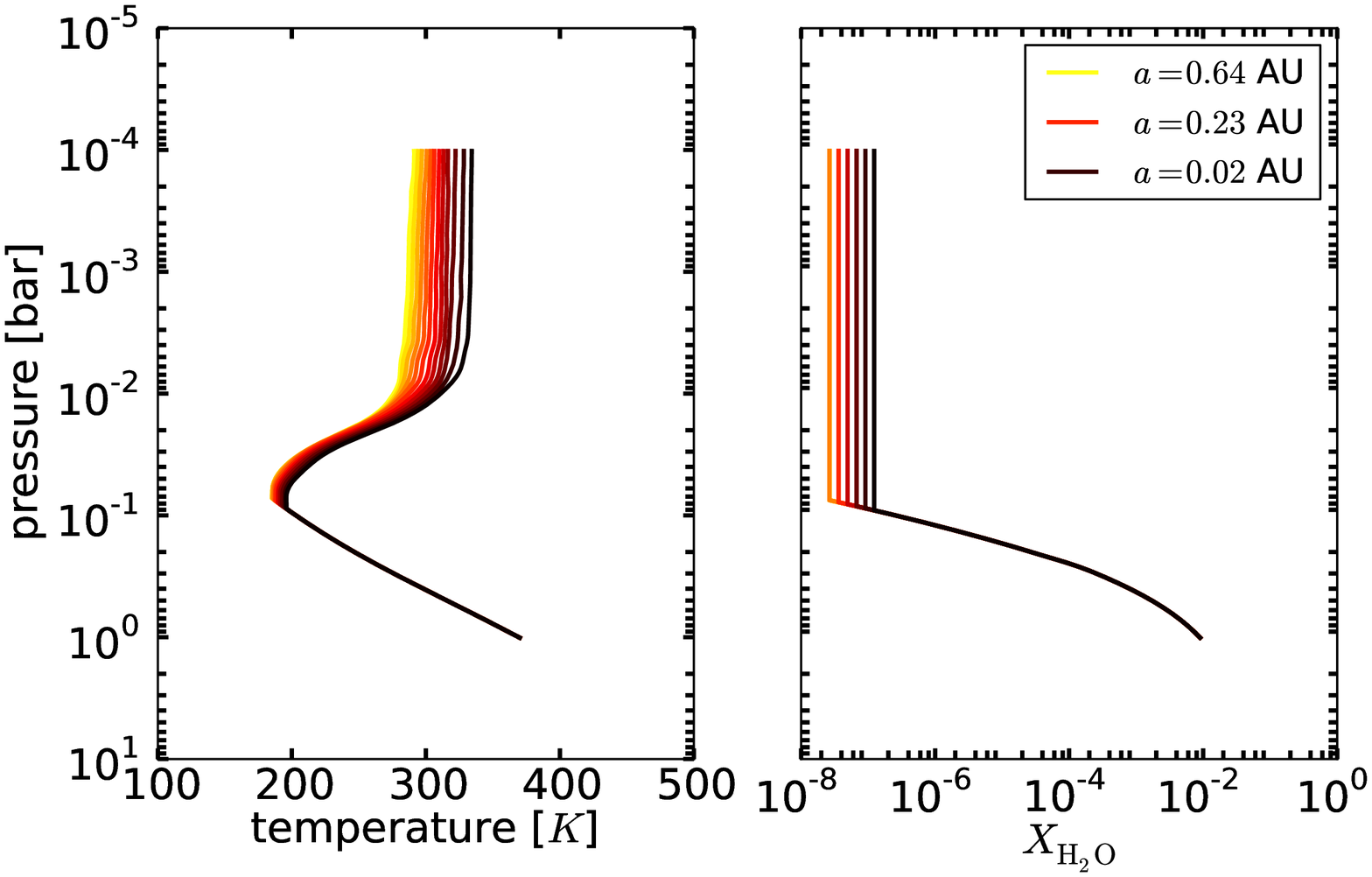}
\caption{\emph{Left:} The temperature and water mixing ratio profiles for a CO$_2$ mixing ratio of 10$^{-5}$ with 1 bar surface pressure, 1\% relative humidity. \emph{Right:} The temperature-pressure and water mixing ratio profiles for a CO$_2$ mixing ratio of 10$^{-1}$, other atmospheric parameters are unchanged. Higher levels of CO$_2$ results in colder tropopause temperatures, and dryer stratospheres at the inner edge. The color coding is identical as in Fig. \ref{fig:relhum_prof_0.01}. \label{fig:CO2_profiles}}
\end{figure*}

\begin{figure}
\centering
\includegraphics[width=0.49\textwidth]{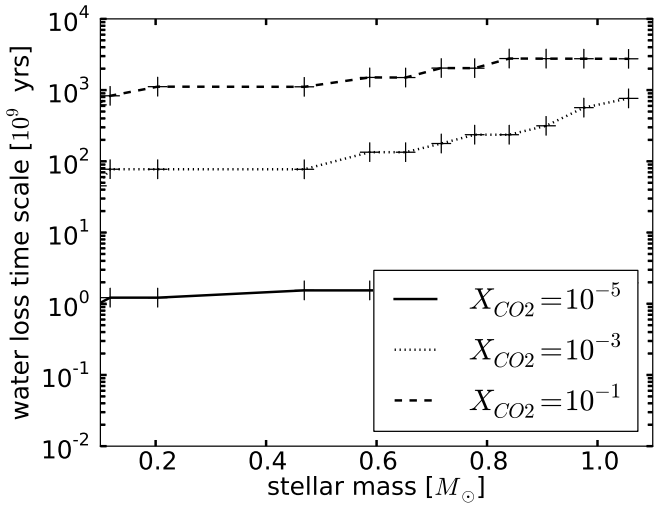}
\caption{The water loss timescale as a function of stellar mass at the inner edge for various CO$_2$ mixing ratios. The relative humidity is 1\% and the surface pressure is 1 bar (for other parameters, see Table \ref{table:para}). As the stratospheric water levels are higher for low CO$_2$ mixing ratios (see Fig. \ref{fig:CO2_profiles}), the water loss timescale becomes shorter. If the CO$_2$ mixing ratio is less than $10^{-4}$, the water loss timescale becomes shorter than 10 billion years. \label{fig:CO2_ocean_loss}}
\end{figure}

The inner edge distance of planets with CO$_2$-rich atmospheres is large. The already knownÊreason is that CO$_2$ is a potent greenhouse gas, thus the outgoing thermal emission decreases with higher CO$_2$ levels. For example, the $OTE$ is 800 W/m$^2$ and 600 W/m$^2$ for $X_{\mathrm{CO_2}} = 10^{-5}$ and $10^{-1}$, respectively. However, the albedo remains largely unaffected. As a result, the inner edge distance is 0.54 AU, and 0.64 AU for CO$_2$ mixing ratios of 10$^{-5}$ and 10$^{-1}$ around a solar-like star.

The tropopause and stratosphere of exoplanets at the inner edge with low levels of atmospheric CO$_2$ are warm (see Fig. \ref{fig:CO2_profiles}). This effect is explained by the variation of the inner edge distance with the CO$_2$ mixing ratio, and by the radiative properties of CO$_2$. Close-in planets are strongly illuminated, therefore the tropopause and the stratosphere are warm. For example, the tropopause temperature is 240 K and 180 K for CO$_2$ levels of 10$^{-5}$ and 10$^{-1}$. CO$_2$ also acts as a coolant in the stratosphere because it emits radiation that is lost to space, and in opposition to water vapor, it does not absorb in optical wavelengths \citep{Clough1995}. The more CO$_2$ the stratosphere has, the cooler it will be.

Close-in planets with low levels of CO$_2$ might quickly lose their surface waters and become unhabitable. The water loss timescale as a function of CO$_2$ mixing ratio is shown in Fig. \ref{fig:CO2_ocean_loss}. For a fixed relative humidity, the tropopause temperature determines the water-loss timescale (see Sect. \ref{subsec:iniprof}). As discussed in the previous paragraph, the troposphere is cold, if the atmosphere is CO$_2$-rich at the inner edge. Therefore, the atmospheric CO$_2$ level has an indirect influence on the water loss timescale, because it regulates the tropopause temperature. This result has been independently verified studied in detail by \citep{Wordsworth2013}.

We note that for a fixed semi-major axis, the CO$_2$ mixing ratio has a similar (but somewhat smaller) effect on the water loss timescale. Radiative-convective models of Earth's atmosphere show that doubling the atmospheric CO$_2$ warms the surface but cools the tropopause \citep{Schlesinger1987}. The result is that the stratospheric water mixing ratio reduces, and the water loss timescale is prolonged. 

\subsection{Intermediate surface pressure necessary for habitability on dry planets}
The impact of surface pressure on the inner edge distance is complex. 

\begin{figure}
\centering
\includegraphics[width=0.49\textwidth]{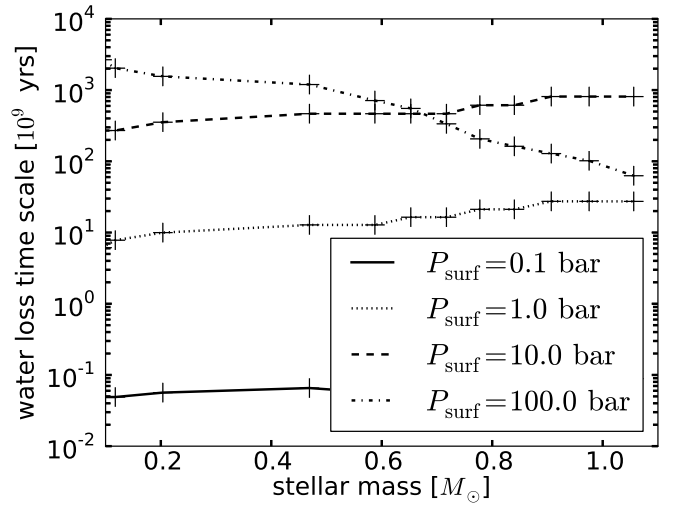}
\caption{The water loss timescale as a function of stellar mass at the inner edge for various surface pressures (see Table \ref{table:para} for other parameters). The water loss timescale is above 10 billion years, if the surface pressure is larger than 1 bar. \label{fig:Psurf_oceanloss}}
\end{figure}

\begin{figure*}
\centering
\includegraphics[width=0.49\textwidth]{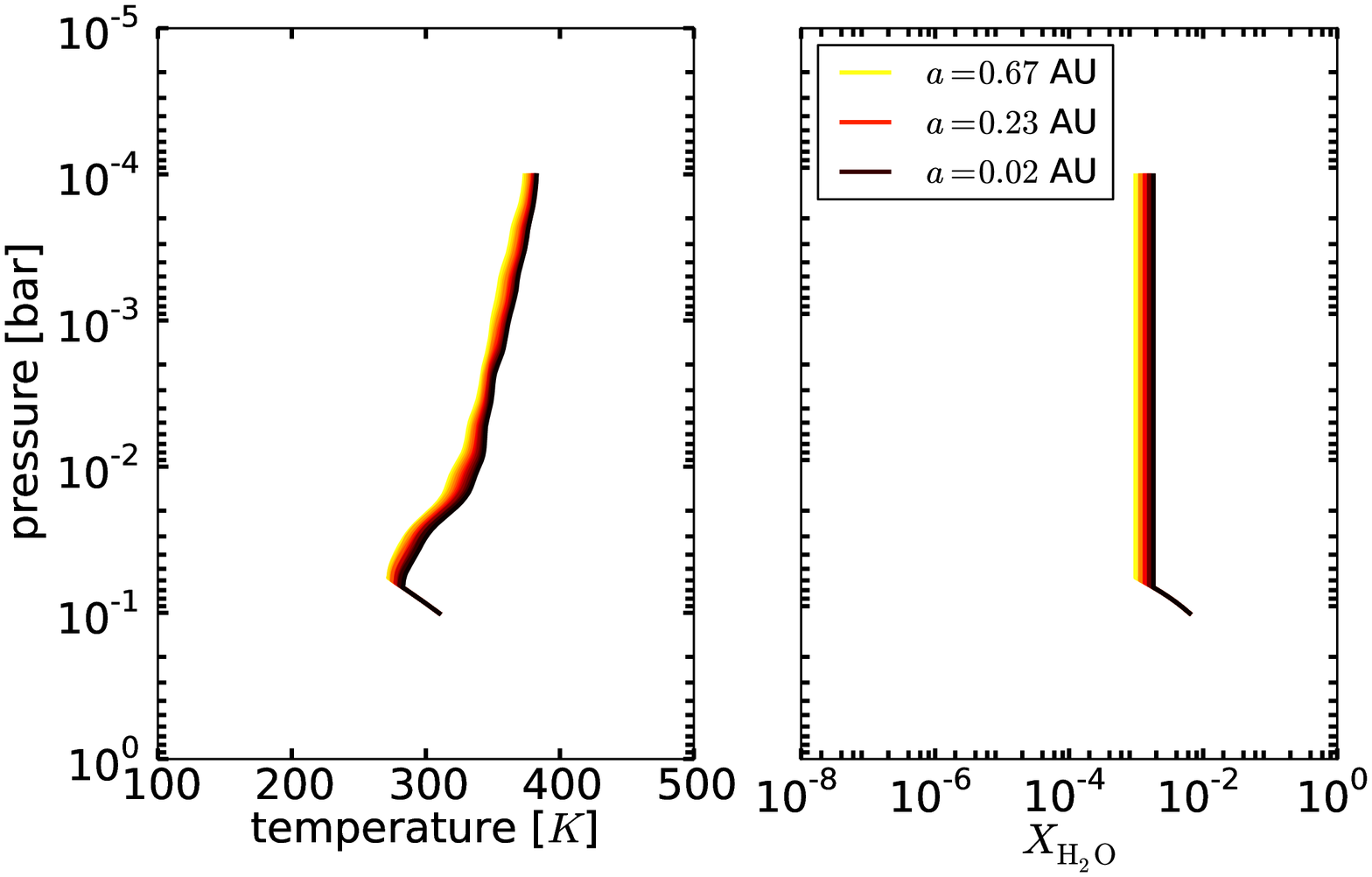}
\includegraphics[width=0.49\textwidth]{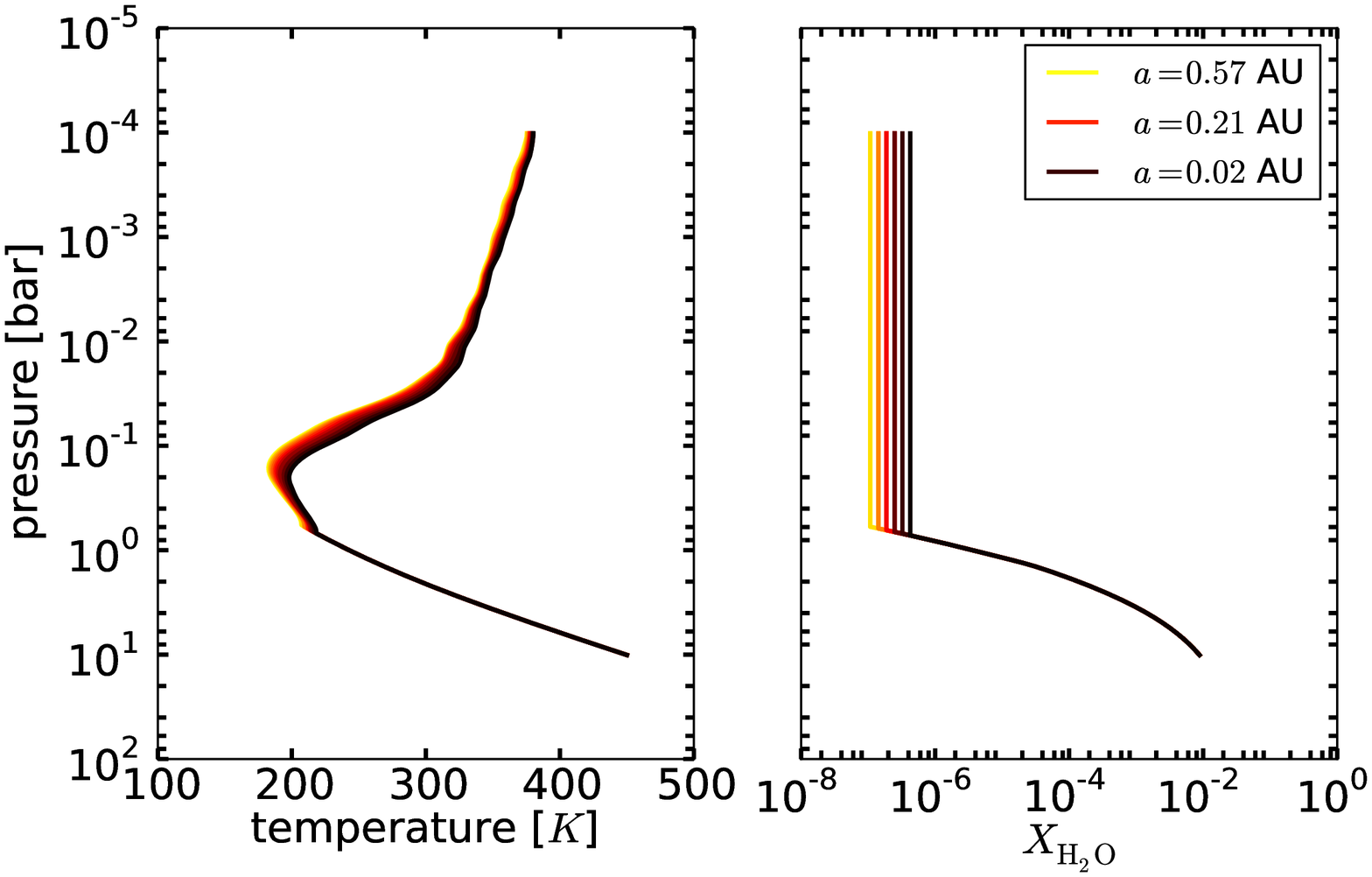}
\caption{\emph{Left:} The temperature and water mixing ratio profiles for an atmosphere with 0.1 bar surface pressure. Other parameters are given in Table \ref{table:para}. \emph{Right:} Atmospheric profiles for a hot desert world with 10 bar surface pressure. The tropopause is much colder in this case. The color coding is identical as in Fig. \ref{fig:relhum_prof_0.01}.\label{fig:Psurf_profiles}}
\end{figure*}

Low surface pressures ($\sim$0.1 bar) are a problem because the water loss timescale is short and complex life might not have enough time to evolve (see Fig. \ref{fig:Psurf_oceanloss}). The water loss timescale of a hot desert world is only 0.1 billion years if the surface pressure is 0.1 bar, this is because the troposphere is confined to low altitudes, and the tropopause temperature only slightly differs from the surface temperature (see Fig. \ref{fig:Psurf_profiles}a). Therefore the stratosphere is water rich and the water loss timescale is short. At large surface pressures, the troposphere is extended, the tropopause is cold (see Fig. \ref{fig:Psurf_profiles}b), and the water loss timescale is always in excess of 10 billion years (see Fig. \ref{fig:Psurf_oceanloss}).

High surface pressures (100 bars) on slow rotators are problematic for habitability. The liquid water cycle cannot plausibly operate because the day-night side temperature difference is too low to initiate condensation. The rotation period of the exoplanet is an important factor influencing the water cycle because it prescribes the dominant circulation mode that redistributes heat. On slow rotators, heat from the day side is transported by zonal flows to the night side (see Sect. \ref{sec:slowrot}). If the surface pressure is 100 bars, the radiative cooling timescale of the atmosphere is long compared to the advection timescale. That is to say, an atmospheric parcel travels across the night side of the planet before it can cool down to the dew point temperature. To give a more qualitative argument, we calculate the surface zonal wind speed necessary to initiate condensation on the night side (Eq. \ref{eq:zonalwind}). The zonal wind speed is on the order of 1000 m/s, 100 m/s, 10 m/s, and 1 m/s for $P_{\mathrm{surf}}$ = 0.1, 1, 10, and 100 bars, respectively. The inverse correlation between surface pressure and zonal wind speed is due to the long radiative cooling timescale at large pressures. The necessary zonal wind speed on exoplanets with 100 bars of surface pressure is too low, therefore, such atmospheres cannot plausibly cool down to the dew point temperature on the night side, and water condensation does not occur. On fast rotators like Earth, baroclinic instability transports heat from the equatorial regions to the poles (see Sect. \ref{sec:fastrot}). We find that the baroclinic diffusion coefficient required to initiate condensation around the poles (Eq. \ref{eq:diff}) is on the order of 10$^{10}$ - 10$^{11}$ m$^2$/s for all surface pressures considered. These values are 4-5 orders of magnitude larger than the baroclinic diffusion coefficient on Earth. Therefore large equator to pole temperature differences and liquid precipitation are plausible on fast rotators at all surface pressures. The baroclinic diffusion coefficient is large because the $OTE$ changes rapidly as a function of surface temperature in dry and hot atmospheres (parameter B in Eq. \ref{eq:diff}).  

\begin{figure}
\centering
\includegraphics[width=0.49\textwidth]{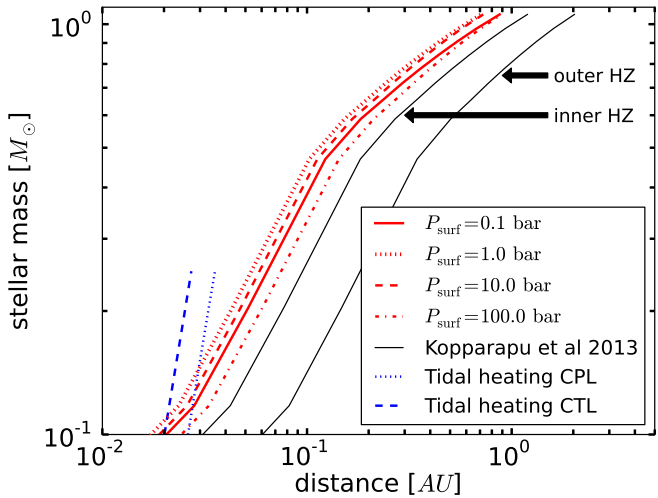}
\caption{The inner edge of the Habitable Zone for various surface pressures. The inner edge limit is minimized for a surface pressure of 1 bar. The Habitable Zone limits of \cite{Kopparapu2013} are indicated with black solid lines as a comparison. The tidal heating limits are indicated with blue lines (for details see the caption of Fig. \ref{fig:relhum_HZlim}). \label{fig:Psurf_HZlim}}
\end{figure}

\begin{figure*}
\centering
\includegraphics[width=0.49\textwidth]{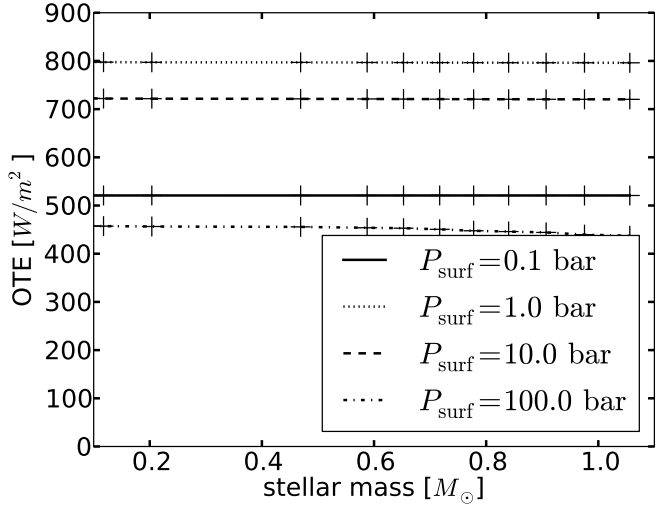}
\includegraphics[width=0.49\textwidth]{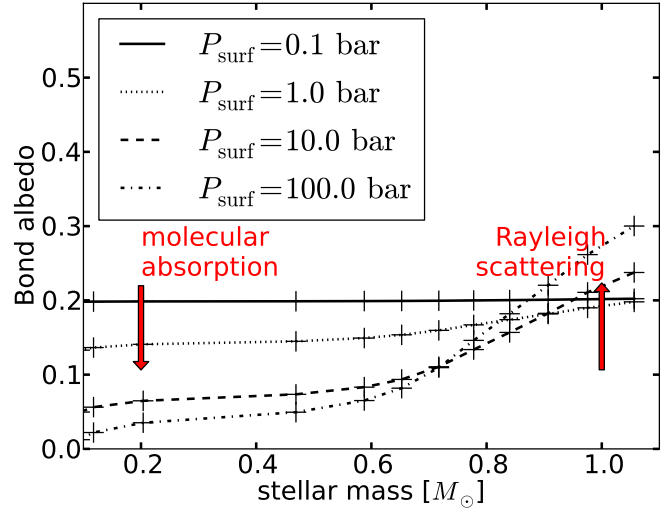}
\caption{\emph{Left:} The outgoing thermal flux for various surface pressures. \emph{Right:} The Bond albedo of exoplanets for various surface pressures. The x axis shows the mass of the host star. \label{fig:Psurf_OTE_alb}}
\end{figure*}

A habitable planet with the smallest semi-major axis should have a surface pressure around 1 bar for stars less massive than 1.1 M$_\odot$ (Fig. \ref{fig:Psurf_HZlim}). The inner edge distance is a non-linear function of the surface pressure and the host star properties, because both the $OTE$ and the albedo show complex features (Fig. \ref{fig:Psurf_OTE_alb}). The $OTE$ is maximized if the surface pressure is 1 bar. However, water boils at low temperatures, if the pressure is reduced. Therefore the surface temperature and thus the $OTE$ are small at low pressures. Although the boiling point and the surface temperature are high at large pressures, the optically thick atmosphere reduces the $OTE$ at large pressures (see Fig. \ref{fig:Psurf_OTE_alb}a). The albedo of low pressure exoplanets is very close to the surface albedo. On the other hand, exoplanets with large surface pressures (100 bars) simultaneously have the lowest albedo around low mass stars due to greenhouse gas absorption; and they also have the highest albedo around massive stars due to Rayleigh scattering (Fig. \ref{fig:Psurf_OTE_alb}b). 

An exoplanet's surface pressure is actually difficult to ascertain whether it be a priori from theory or from observations. The surface pressure of exoplanets is influenced by processes such as accretion from the protoplanetary disk, atmospheric escape, atmosphere accretion and/or erosion during impacts, and outgassing. In fact, the most challenging problem in proving that an exoplanet is habitable will be to observationally constrain its surface pressure. Even if the atmosphere is cloud-free, the atmosphere could be so optically thick that the surface remains hidden and unconstrained. 

\subsection{Surface gravity weakly influences the inner edge}
\begin{figure}
\centering
\includegraphics[width=0.49\textwidth]{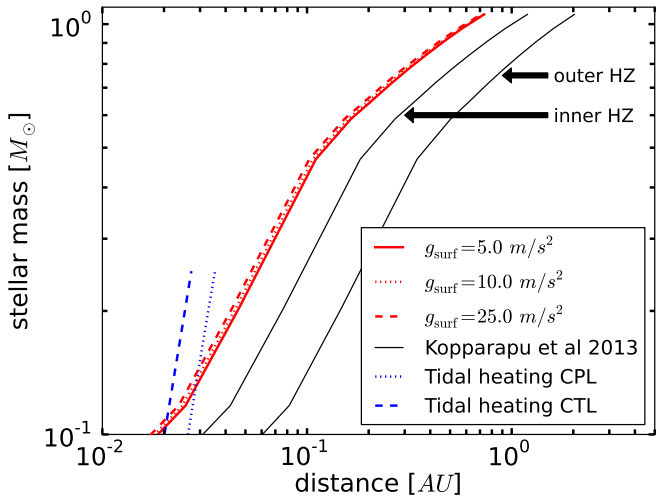}
\caption{The inner edge of the Habitable Zone for various surface gravities. The x axis is the planet-star separation, the y axis shows the mass of the host star. Surface gravity has a small effect on the inner edge distance for the atmospheric configuration adapted. The inner edge is somewhat closer to the host star, if the surface gravity is small. The Habitable Zone limits of \cite{Kopparapu2013} are shown with black solid lines. The tidal heating limits are indicated with blue lines (for details see the caption of Fig. \ref{fig:relhum_HZlim}). \label{fig:gsurf_HZlim}}
\end{figure}

\begin{figure}
\centering
\includegraphics[width=0.49\textwidth]{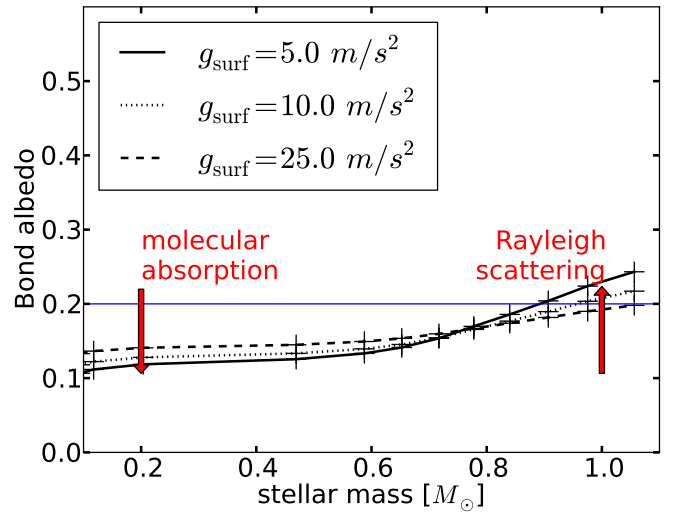}
\caption{The Bond albedo of exoplanets for various surface gravities (other parameters given in Table \ref{table:para}). The blue solid line indicates the surface albedo of 0.2 as a reference. The outgoing thermal emission (not shown) is 700, 745, and 800 W/m$^2$ for surface gravities of 5, 10, and 25 m/s$^2$, respectively. As illustrated in Fig. \ref{fig:gsurf_HZlim}, surface gravity has a modest influence on the inner edge distance. \label{fig:gsurf_albedo}}
\end{figure}

The effect of surface gravity on the inner edge distance is small compared to other factors discussed previously (see Fig. \ref{fig:gsurf_HZlim}). Surface gravity influences the pressure scale height of the atmosphere and thus the column density of gases for a fixed surface pressure. The atmosphere of a Mars-sized small planet is expanded and thus the column density of all gases is raised. This has implications for the $OTE$ as well as the planetary albedo. Increasing the greenhouse gas column density reduces the $OTE$ from 800 W/m$^2$ to 700 W/m$^2$ for planets with surface gravities of 25 m/s$^2$ and 5 m/s$^2$, respectively. The albedo of a Mars-sized small planet shows the largest deviation from the surface albedo (see Fig. \ref{fig:gsurf_albedo}). The albedo around low mass stars is the lowest for small planets, because the greenhouse gas column density is large and the incoming stellar light is absorbed. At the same time, the albedo around massive stars is also the largest for small planets because Rayleigh scattering is more efficient due to the large column density of N$_2$. 

Although the differences are small, the inner edge is slightly closer to the host star for super-Earth planets (see Fig. 8b). The distance around a solar-like star is 0.57 AU, 0.56 AU, and 0.55 AU for surface gravities of 5, 10, and 25 m/s$^2$. As the inner edge distance changes rather moderately with planet mass and size, it is a good approximation to use one common inner edge limit for all super-Earth and Earth-like planets. 

\subsection{Reflective surfaces push the inner edge close to the star}
\begin{figure}
\centering
\includegraphics[width=0.49\textwidth]{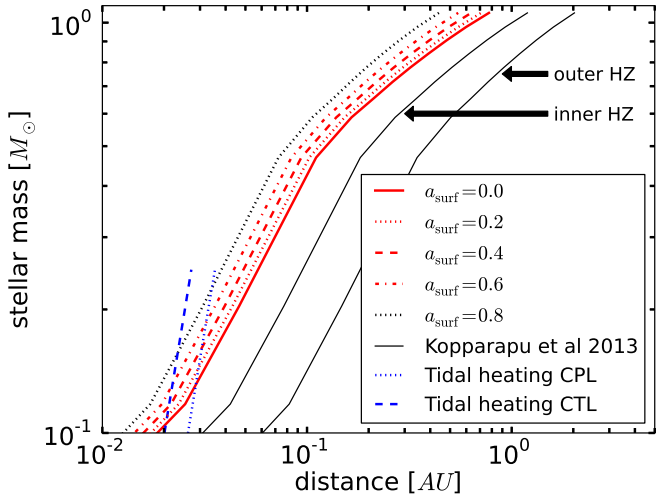}
\caption{The inner edge of the Habitable Zone for various surface albedos. The x axis is the planet-star separation, the y axis shows the mass of the host star. The inner edge is at 0.38 AU around a solar-like star (inside the orbit of Mercury), if the surface albedo is 0.8. The Habitable Zone limits of \cite{Kopparapu2013} are shown with black solid lines. The tidal heating limits are indicated with blue lines (for details see the caption of Fig. \ref{fig:relhum_HZlim}). \label{fig:asurf_HZlim}}
\end{figure}

If the surface pressure is moderate ($P_{\mathrm{surf}} \leq 10$ bars), the atmosphere is optically thin in optical wavelengths. Therefore, a significant fraction of the stellar radiation reaches the surface, and the surface albedo has a strong influence on the inner edge distance (Fig. \ref{fig:asurf_HZlim}). If the surface albedo is 0.8 (a hypothetical, possibly unrealisticly high value), the inner edge of the Habitable Zone is at 0.38 AU around a solar-like star (within the orbit of Mercury). Such a high surface albedo pushes the inner edge to the extremes and it remains to be seem whether we find any planet with a highly reflective surface. In the following we adopt two inner edge limits: the nominal atmosphere with properties outlined in Table \ref{table:para}), and one with a surface albedo of 0.8.

\subsection{Application to confirmed/validated super-Earth exoplanets and candidates}
\label{subsec:HZlim}
We approximate our inner edge distances using a simple power function in the form:
\begin{equation}
d_{in} = d_{in\odot}L^s,
\label{eq:fit}
\end{equation}
where $d_{in}$ is the inner edge distance in AU, $L$ is the stellar luminosity in units of solar luminosity (L$_\odot$), $d_{in\odot}$ is the inner edge distance at $L = 1$ L$_\odot$, and $s$ is the slope of the power function. We find that $d_{in\odot} = 0.59$ AU and $s = 0.495$ for our nominal scenario with parameters shown in Table \ref{table:para}. If the surface albedo is 0.8 in the nominal case, $d_{in\odot} = 0.38$ AU and $s = 0.474$. The goodness-of-fit estimates on these parameters (1 $\sigma$) is in the third decimal point.

We show currently confirmed/validated and candidate super-Earth exoplanets in relation to our fundamental inner edge boundary and to the HZ limits of \cite{Kopparapu2013}. We use the Open Exoplanet Catalogue of \cite{Rein2012} as the source of stellar and exoplanetary data for confirmed/validated exoplanets. The NASA Exoplanet Archive is used to access the Kepler exoplanet candidates of the Q1-Q12 data sets \citep{Burke2013}. The stellar and planetary properties of low mass stars are updated according to \cite{Dressing2013}. We consider exoplanet (candidates) that have a mass (or $m\sin i$) less than 10 $M_\oplus$, and/or a radius less than 2.5 $R_\oplus$.

\begin{figure*}
\centering
\includegraphics[width=0.49\textwidth]{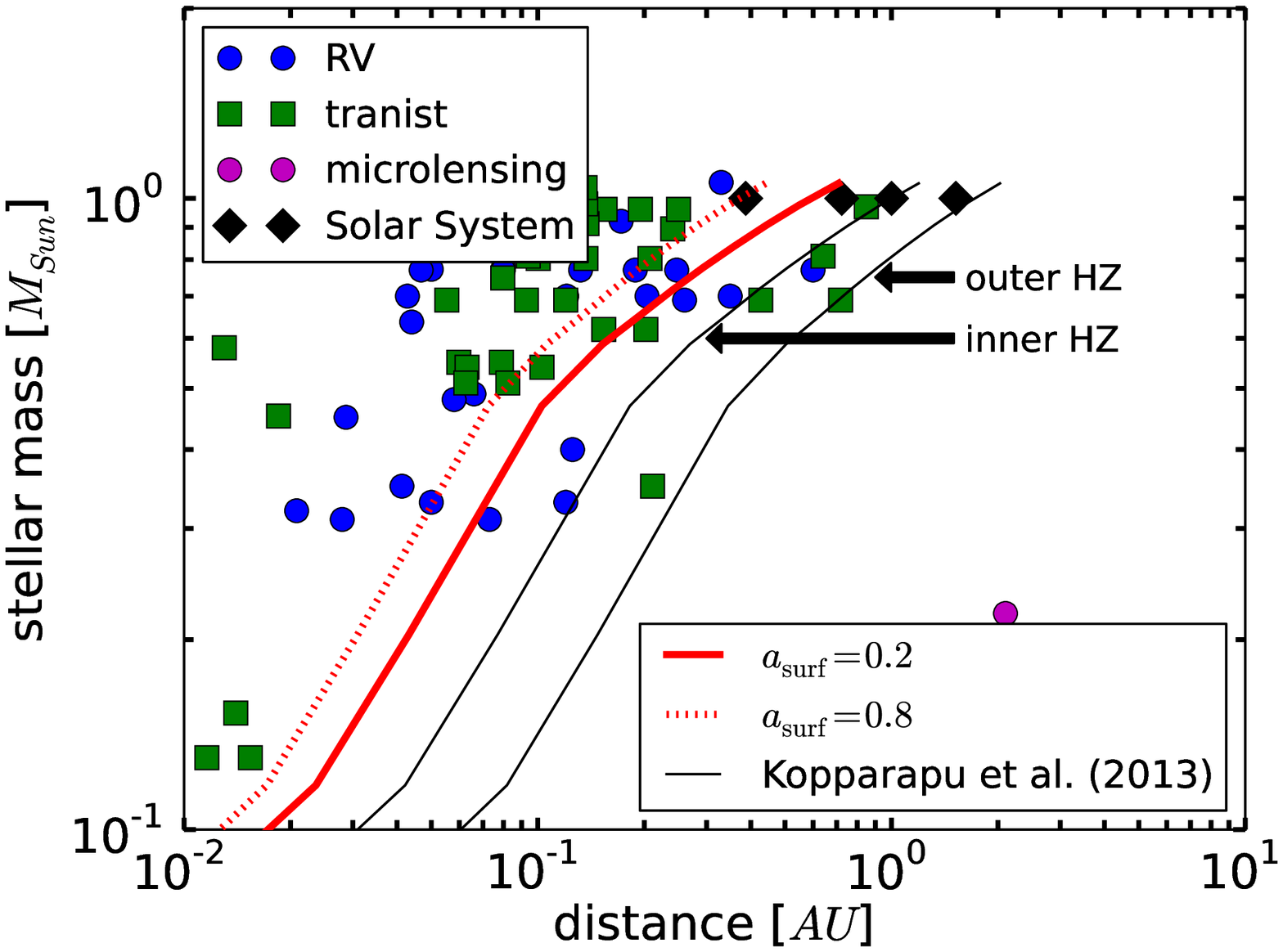}
\includegraphics[width=0.49\textwidth]{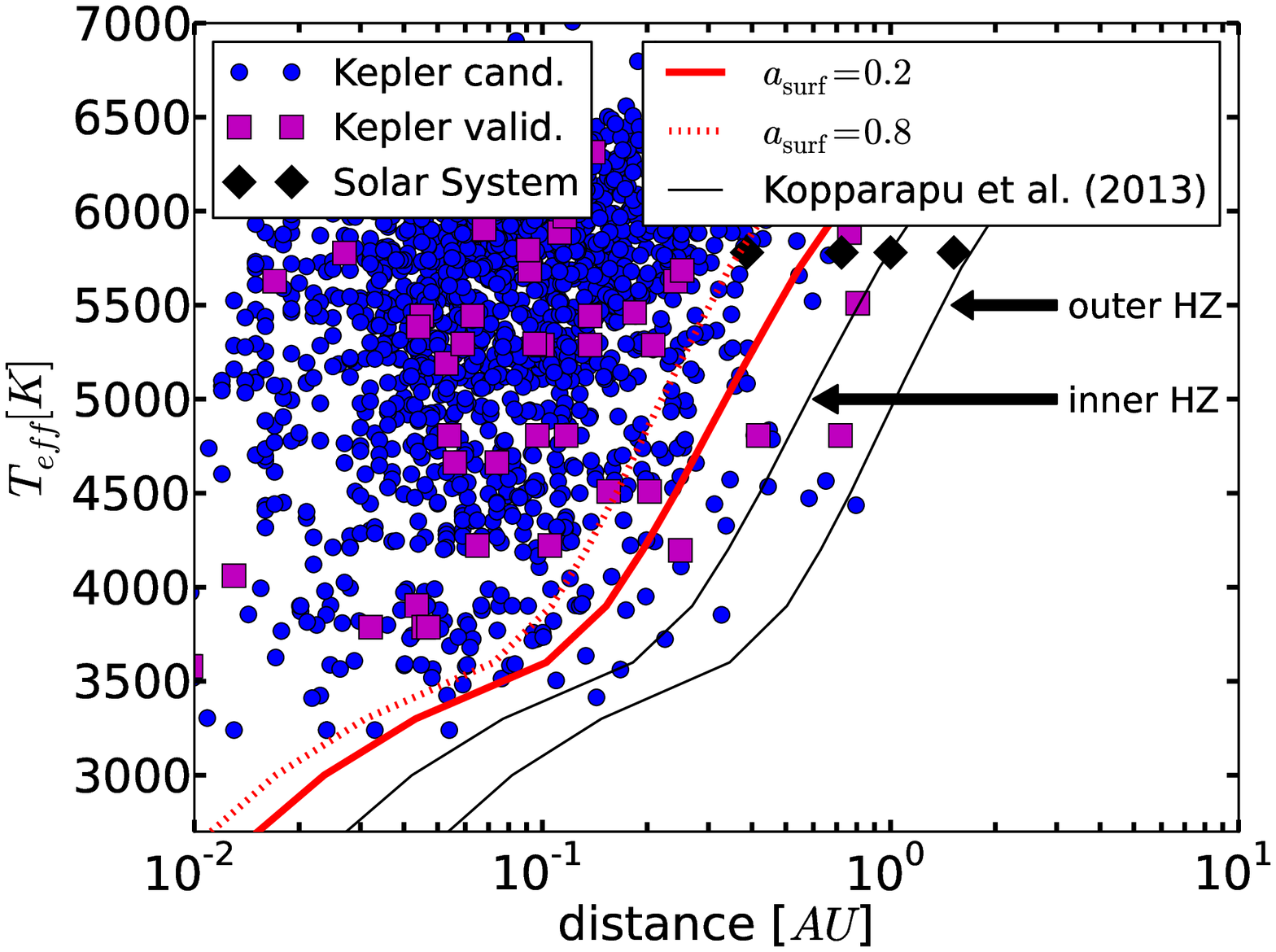}
\caption{\emph{Left:} The semi-major axes of confirmed/validated super-Earth planets ($m\sin i$ less than 10 $M_\oplus$, and/or a radius less than 2.5 $R_\oplus$) as a function of their host stars' mass. The inner edge of the Habitable Zone assuming a surface albedo of 0.2 (solid red line) and 0.8 (dotted red line) are also shown. Exoplanets discovered by various detection methods are included in the figure, as well as the Habitable Zone limits of \cite{Kopparapu2013}, and the positions of the four rocky Solar System planets. \emph{Right:} The semi-major axes of confirmed/validated and candidate Kepler super-Earth planets as a function of the host stars' effective temperatures. The number of potentially habitable exoplanets increases by a factor of 2-3 compared to previous estimates. However, follow-up observations aimed to characterize super-Earth atmospheres are necessary to determine whether these exoplanets are really habitable. \label{fig:HZ}}
\end{figure*}

There are ten radial velocity super-Earths, eight validated transiting exoplanets and one exoplanet discovered by microlensing that could potentially be habitable given suitable atmospheric conditions. The confirmed/validated exoplanets are shown in Fig. \ref{fig:HZ}a. The potentially habitable radial velocity planets with increasing host star mass are Gliese 581c \citep{Udry2007}, Gliese 667Cc \citep{Anglada-Escude2012}, Gliese 163c \citep{Bonfils2013},HD 85512b \citep{Pepe2011}, HD 20794c and d \citep{Pepe2011}, and HD 40307f and g \citep{Tuomi2013}. The transiting potentially habitable exoplanets with increasing stellar mass are Kepler 61b \citep{Ballard2013}, Kepler 52c \citep{Steffen2013}, Kepler 55b and c \citep{Steffen2013}, Kepler 62e and f \citep{Borucki2013}, Kepler 69c \citep{Barclay2013}, and Kepler 22b \citep{Borucki2012}. The microlensing exoplanet OGLE-2005-BLG-390Lb \citep{Beaulieu2006} could also be habitable, if its atmosphere is hydrogen-rich \citep{Pierrehumbert2011a}.

We identify on the order of 100 potentially habitable Kepler candidates that have a radius below 2.5 $R_\oplus$. Although the Kepler mission is not designed to study nearby low mass stars, it did discover a large population of planetary candidates around low mass stars \citep{Dressing2013}. We anticipate that the Transiting Exoplanet Survey Satellite \citep[TESS,][]{Ricker2010} will discover a large number of super-Earth and Earth-sized planets around nearby low mass stars. Due to the proximity of the Habitable Zone to the host star, a large fraction of those planets could potentially be habitable. 

We find that the occurrence rate of terrestrial planets (with radius between 0.5 and 2 $R_\oplus$) in the Habitable Zone for M dwarfs is around 0.7-0.9. In comparison, it is 0.5-0.6 based on the Earth-analog Habitable Zone limits of \cite{Kopparapu2013} \citep{Kopparapu2013a}. The occurrence rate is $0.70\substack{+0.15 \\ -0.09}$ for the nominal scenario with a surface albedo of 0.2, and $0.88\substack{+0.09 \\ -0.05}$ for the high surface albedo case. We apply the method of \cite{Dressing2013} to estimate the occurrence rate and assume that the outer edge of the Habitable Zone is at infinity. One caveat of the estimate is that the occurrence rate estimate is not statistically robust yet because the number of planets detected at large separations is small. KOI 2418.01 contributes 0.1 planet/star to the total occurrence rate and KOI 1686.01 contributes 0.24 planets/star. This means that the true occurrence rate within the HZ might be significantly lower if either KOI 2418.01 or 1686.01 are false positives.

\section{Discussion}
\label{sec:discussion}

\subsection{The formation of desert worlds}
\label{subsec:form}
Where should we look for hot desert worlds? Is it possible to estimate their occurrence rate? What factors play a significant role in forming desert worlds? To answer these questions, we rely on N-body simulations of interacting planetary embryos and planetesimals \citep[e.g.,][]{Morbidelli2000, Chambers2001, Raymond2004, Raymond2006, Raymond2007a}. Such models simulate how planetary embryos accrete water-rich km-sized planetesimals coming past the snow line. 

N-body simulations show that water delivery onto rocky planets is a stochastic process and the water reservoir of exoplanets might be difficult to constrain from models. One general trend that emerges from these simulations is that the water reservoir of the planet correlates with the surface density of icy solids around the snow line. However, as water delivery occurs mostly during the planet formation stage, the surface density of solids might be impossible to observationally constrain for old systems.

Water rich planets close to the host star could naturally end up with small water reservoirs after the formation process \citep{Hamano2013}. It is expected that newly formed rocky planets are in a molten state because they form via giant impacts. If the planet is close to the host star, the surface is unable to cool and solidify efficiently due to the greenhouse effect of the water vapor dominated atmosphere.  As a result, the planet can experience a prolonged run-away greenhouse stage and lose most of its water even if the planet started out with a larger water reservoir than Earth. This process provides a robust way to form dry planets close to the host star.

\subsection{Can dry planets keep their water reservoir?}
Surface water can be removed by the hydration of a planet's crust and potentially upper mantle. On Earth, most hydration occurs along mid-ocean ridges caused by plate tectonics \citep[see e.g.,][and references therein]{Kasting1992}. However, during subduction water is again released through volcanism - possibly stabilizing the amount of surface water on plate tectonics planets. Hence, as a first estimate, plate tectonics planets seem to be capable of keeping small surface oceans.

Planets lacking plate tectonics (stagnant lid planets) are expected to have lowered crustal and mantle hydration rates due to the lack of mid-ocean ridges but it is not clear how effective volcanism returns the absorbed water to the atmosphere. We could assume that rock hydration occurs primarily by, e.g., serpentinization (or by the formation of clays). Both are low temperature processes and do generally not occur above 500 K - 700 K \citep[for serpentinization,][]{Oze2007} and demand large permeabilities and pore spaces for water to penetrate. Hence any mineral hydration strongly decays with depth due to increasing interior temperatures and decreasing permeabilities.

Taking Earth as an example (although Earth has plate tectonics), the current average geotherm of $\sim$500 K - 700 K is reached at maximally $\sim$10-15 km depth. This is assuming a 290 K surface temperature and a 25 K/km continental thermal gradient \citep[e.g.,][]{Turcotte2002}. However, the permeability limits water penetration down to maximally $\sim$5 km \citep{Vance2007}. Assuming the unrealistic case that 5 km on Earth are being completely serpentinized \citep[molar weight of forsterite $\sim$0.14 kg/mol, water 0.018 kg/mol, average forsterite density of $\sim$3300 kg/m$^3$, and water to forsterite molar reaction ratio of 1.5, e.g.,][]{Oze2007} would indeed allow to completely absorb the Earth's ocean after $\sim$5 billion years. However, the assumptions above are highly unrealistic as they assume that the whole crust is serpentinized and that it consists of pure forsterite. As we know from Earth, serpentinization is a highly local and irregular phenomenon, and hence the maximally serpentinized crustal volume is expected to be by orders of magnitude smaller than our worst-case estimate \citep[i.e.,][]{Nazarova1994}.

Moreover, our estimation neglects the return of water by volcanic outgassing and the fact that hot desert planets (especially the more massive ones) are different from Earth: due to larger surface temperatures (and especially for more massive planets) the interior is hotter and hence the maximally hydrated lithospheric volume is smaller than estimated for Earth.

As a first approximation, small oceans seem to be feasible over geologic timescales on plate tectonics but also on stagnant lid planets, though clearly more work is needed to understand the geophysical conditions of hot desert worlds, especially if they do not facilitate plate tectonics.

\subsection{Observables}
\label{subsec:obs}
A conclusive evidence of habitability would be to detect liquid water on the surface of a rocky exoplanet. Transmission spectroscopy can only infer habitability because it probes the atmosphere and not the surface. In other words, it will be a necessary but not sufficient proof of habitability to identify water vapor in the atmosphere of a rocky exoplanet using transmission spectroscopy.

\begin{figure}
\centering
\includegraphics[width=0.49\textwidth]{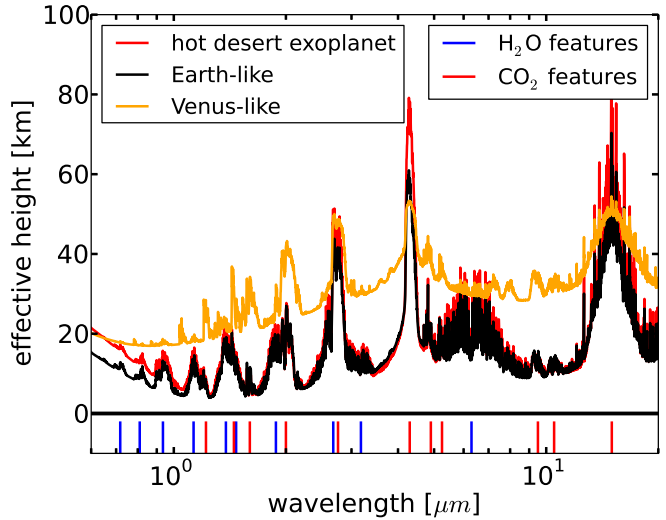}
\caption{The transmission spectra of a hot desert world with 10 m/s$^2$ surface gravity (other parameters in Table \ref{table:para}), Venus (with the 0 km height moved to 1 bar from the surface), and an Earth-like planet (with only N$_2$, water and CO$_2$ in the atmosphere). The x axis is the wavelength, the y axis is the effective height of the atmosphere. Medium resolution spectra obtained with the JWST will enable us to distinguish a hot desert world from a Venus-like atmosphere. The effective height of an Earth-like planet and a hot desert world look similar. Therefore, an instrument designed to characterize biosignature gases in an Earth-like atmosphere can also identify similarly abundant biosignature gases in the atmosphere of a hot desert world. \label{fig:trans_height}}
\end{figure}

We show the effective height as a function of wavelength \citep{Kaltenegger2009} for three types of atmospheres in Fig. \ref{fig:trans_height}: a hot desert world with 10 m/s$^2$ surface gravity (see Table \ref{table:para} for other parameters); the effective height of a CO$_2$ dominated Venus-like atmosphere with a water mixing ratio of $3\times10^{-5}$; and the effective height of an Earth-like atmosphere with N$_2$/CO$_2$/H$_2$O (other gases are removed for easier comparison). The spectra are computed at high resolution ($\lambda / \delta \lambda = 10^5$) and binned down to a resolution of 1000, which is comparable to the medium resolution of NIRSPEC on the \textit{James Webb Space Telescope} \citep{Bagnasco2007}. The illustrated transmission signals are noise-free, but a typical signal-to-noise ratio is expected to be $\sim$10 for water features at 2 micron for systems out to 12 pc, assuming a total observing time of 200 hours \citep{Deming2009}. 

The transmission signal of an Earth-like and hot desert exoplanets are similar, because the stratosphere of these atmosphere types are similar and transmission spectroscopy mainly probes the stratosphere. Water vapor is abundant in the troposphere close to the surface, if a surface reservoir is present. However, transmission spectroscopy typically constrains the stratospheric compositional information only, because the surface region is expected to be optically thick at all wavelengths, and thus the transmission signal is, for the most part, unaffected by the troposphere. This is a problem, because surface properties such as surface pressure and water reservoir remain unconstrained with observations. This point is illustrated in Fig. \ref{fig:trans_height} because the water features of a hot desert world and an Earth-like planet are similar, although the surface relative humidity differs by almost two orders of magnitude. Furthermore, the effective height is above 5 km for both cases, thus the surface regions are unconstrained. The largest difference in the transmission signal is at wavelengths shorter than 1 micron because this part of the spectrum is influenced by the pressure scale height of the atmosphere. The pressure scale height is 
\begin{equation}
H_p = \frac{kT}{Mg_{\mathrm{surf}}},
\label{eq:presscale}
\end{equation}
where $k$ is the Boltzmann constant, $T$ is the temperature of the atmosphere, and $M$ is the mean molecular mass. The mean molecular mass is similar in the two cases, the surface gravities are identical. However, the temperatures on a hot desert world are larger than on Earth. The average surface temperature of the hot desert world is 370 K in contrast to the 290 K surface temperature of Earth. The amplitude of the molecular features differs at all wavelengths. This difference is also explained by the large pressure scale height on hot desert worlds.

The transmission signal of a Venus-like atmosphere is significantly different from the expected signal of a hot desert exoplanet. Therefore, it might be possible to identify habitable planets with confidence. If the stratospheric mixing ratios of CO$_2$ and water are retrievable from the transmission spectrum, it is possible to distinguish a CO$_2$-dominated atmosphere from an N$_2$-dominated atmosphere \citep{Benneke2012}. Variations in the effective height are modest in a CO$_2$ dominated atmosphere for two reasons. First, the CO$_2$ absorption lines are wide and the atmosphere absorbs significantly even in between band center regions. Second, the scale height of a CO$_2$ dominated atmosphere is small as the molecular weight of CO$_2$ is larger than the molecular weight of N$_2$. However, if the CO$_2$ and water mixing ratios are small, the variations in effective height are large (see Fig. \ref{fig:trans_height}). Measuring the relative effective heights of band centers and window regions is a way to distinguish CO$_2$-dominated atmospheres from Earth-like and hot desert world atmospheres.

\begin{figure}
\centering
\includegraphics[width=0.49\textwidth]{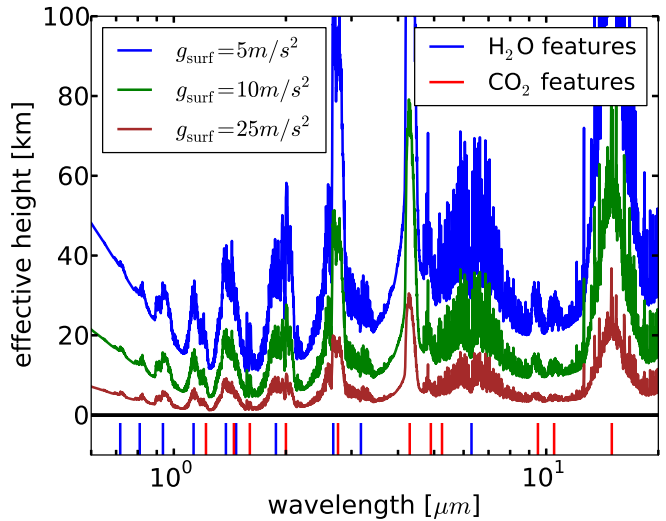}
\caption{The transmission spectra of hot desert worlds with various surface gravities (for other parameters, see Table \ref{table:para}). The surface gravity influences the pressure scale height of the atmosphere. The molecular features are easier to detect if the exoplanet's surface gravity is small, because the pressure scale height is large. \label{fig:trans_height_gsurf}}
\end{figure}

We emphasize the previously known result that the transmission signal of low surface gravity planets is easier to detect than the signal of high surface gravity planets because of a larger scale height (and thus larger variations in the transmission signal) for the former. We illustrate the influence of surface gravity on the effective height for hot desert worlds in Fig. \ref{fig:trans_height_gsurf}. 

\subsection{Model uncertainties}
\label{subsec:uncert}

\subsubsection{Clouds}
\label{subsec:clouds}
The effects of water clouds on the inner edge distance of desert worlds is presumably small, because we do not expect tenuous water clouds covering a large fraction of the surface on dry worlds. However, there are some cases where clouds could influence the inner edge distance. Most notable are clouds forming in the convective equatorial regions of the Hadley cell. Cloud formation is initiated by condensation, thus a large surface-tropopause temperature difference is necessary to form convective clouds. The temperature difference is large if the CO$_2$ mixing ratio and/or the surface pressure are large. Even if clouds do form, it is expected to happen high in the troposphere \citep{Zsom2012}. Therefore, the clouds might be thinner on hot desert worlds than on Earth. Furthermore, it is difficult to a priori estimate whether such clouds would reduce or increase the inner edge distance because the albedo effect and the warming effect of high water clouds are typically of similar magnitude \citep{Goldblatt2011}. 

Clouds forming on the night side of slow rotators could also influence the inner edge distance, if such clouds extend to the day side. If zonal winds shepherd clouds to the day side, and the clouds survive to the vicinity of the sub-stellar point, the albedo effect of clouds could push the inner edge closer to the host star. Clouds form predominantly at the poles on fast rotators assuming that their rotation axis is close to perpendicular to the orbital plane, thus their effect on the inner edge distance is limited due to the large zenith angle of incoming stellar radiation. 

\subsubsection{1D vs. 3D}
The main assumption in our approach is that the mixing ratio of atmospheric constituents and the temperature-pressure profile is described by 1D vertical globally averaged quantities. If the mixing ratio of a chemical is influenced by processes operating on short timescales relative to atmospheric mixing, spatial variations in mixing ratio throughout the atmosphere are expected to be large. For example, if the atmospheric CO$_2$ is regulated by the carbon-silicate cycle, the assumption of global averaging is valid, because the carbon-silicate cycle operates on a long timescale compared to atmospheric mixing \citep{Walker1981}. However, the mixing ratio of H$_2$O could show large horizontal variations because the surface waters are expected to be confined to a small portion of the surface on hot desert worlds, and precipitation is not anticipated on the day side. Such mixing ratio variations could have important implications for planetary climate and the inner edge distance because regions with relative humidities below average act as global coolants. Two examples are the dry downwelling part of the Hadley cell or the Foehn wind. The optical depth in these columns is reduced and more infrared radiation escapes to space thus cooling the planet. Areas with relative humidities above average are expected where precipitation occurs (night side or poles). The enhanced greenhouse effect in these areas locally increases the cooling timescale (Eq. \ref{eq:tau_rad}), thus reducing the overall horizontal temperature gradients. Multidimensional coupled radiative and circulation models are necessary to study these effects on the climate in greater depth.

\subsubsection{Relative humidity profile}

Another crucial assumption is the constant tropospheric relative humidity profile. The relative humidity profile determines the column density of water vapor in the troposphere and thus the radiative properties of the atmosphere. The relative humidity profile on Earth decreases with altitude being 70\% on the surface and a couple of percentage at the tropopause \citep{Manabe1967}. The relative humidity profile is determined by the water cycle: surface evaporation, atmospheric circulation, condensation, and precipitation. It is beyond the scope of this paper to study such effects but we plan to investigate this problem in the future.

\section{Summary and Conclusions}
\label{sec:sum}
We studied a wide range of atmospheric configurations that provide habitable surface conditions while minimizing the semi-major axis of the exoplanet. The inner edge of the habitable zone is pushed close to the host star if the greenhouse effect is reduced (low relative humidity) and the surface albedo is increased. We argue that if precipitation occurs predominantly in liquid form, the habitable region on the planet's surface remains extended. The requirement for liquid precipitation provides a lower limit on relative humidity, which we estimate to be 1\% for a wide range of surface temperatures and pressures. We studied the atmospheric circulation requirements to initiate condensation and precipitation either on the night side or at the poles of exoplanets. The water loss timescale was also estimated to assess the life-time of the surface water reservoir. Note that we focused exclusively on the inner edge of the Habitable Zone because the outer edge extends well beyond the N$_2$/H$_2$O/CO$_2$ Habitable Zone limits of \cite{Kasting1993,Kopparapu2013} for H$_2$-He dominated atmospheres \citep{Pierrehumbert2011a}, and it can extend to infinity, if the exoplanet has sufficient internal heat sources and a H$_2$-rich atmosphere \citep{Stevenson1999}.

Our main results are summarized here:
\begin{itemize}
\item We estimate that the fundamental inner edge of the Habitable Zone can be as close as 0.38 AU around a solar-like star (within the orbit of Mercury), if the relative humidity of the planetary atmosphere is low (1\%) and the surface albedo is large (0.8). If the surface albedo is moderate (0.2), the inner edge is at 0.59 AU for the same relative humidity. In both cases, we assume a surface pressure of 1 bar, and $10^{-4}$ CO$_2$ mixing ratio.
\item A lower limit on the atmospheric CO$_2$ level exists because the CO$_2$ mixing ratio influences the water loss time scale. We find that the water loss timescale is longer than 10 billion years. Thus life can plausibly evolve on hot desert planets, if the atmospheric CO$_2$ level is above $10^{-4}$ and the surface pressure is more than 1 bar. 
\item If the surface pressure on slow rotator planets is on the order of 1-10 bars, it is plausible that the night side cools down to the dew point temperature and precipitation is triggered. For large surface pressures (e.g., 100 bars), the heat capacity of the atmosphere is large and it prevents strong cooling and large day-night temperature differences.
\item Liquid water precipitation is plausible at the poles of fast rotators, because the minimum required baroclinic diffusion coefficient to initiate precipitation is large for all atmospheric scenarios considered here.
\item The number of potentially habitable confirmed/validated exoplanets significantly increases. The potentially habitable super-Earth exoplanets are \object{Gliese 581c}, \object{Gliese 667Cc}, \object{Gliese 163c}, \object{HD 85512b}, \object{HD 20794c} and \object{HD 20794d}, \object{HD 40307f} and \object{HD 40307g}, \object{Kepler 61b}, \object{Kepler 52c}, \object{Kepler 55b} and \object{Kepler 55c}, \object{Kepler 62e} and \object{Kepler 62f}, \object{Kepler 69c}, and \object{Kepler 22b}. The microlensing exoplanet \object{OGLE-2005-BLG-390Lb} could also be habitable, if its atmosphere is hydrogen-rich.
\item We provide empirical fitting formulas for our inner edge limits in a form of a power law: $d_{in} = d_{in\odot}L^s$, where $d_{in}$ is the inner edge distance in AU, $L$ is the stellar luminosity in units of solar luminosity (L$_\odot$), $d_{in\odot}$ is the inner edge distance at $L = 1$ L$_\odot$, and $s$ is the slope of the power function. We find that $d_{in\odot} = 0.59$ AU and $s = 0.495$ for a surface albedo of 0.2 and 1\% relative humidity (see Table \ref{table:para} for other parameters). If the surface albedo is 0.8, $d_{in\odot} = 0.38$ AU and $s = 0.474$ for the same relative humidity.
\item The occurrence rate of terrestrial planets (with radius between 0.5 and 2 $R_\oplus$) in the Habitable Zone for M dwarfs could be as large as 0.7-0.9 using our inner edge estimates.
\item We find that the transmission spectra of Earth-like and hot desert exoplanets are similar. Therefore an instrument designed to characterize the atmosphere of an Earth-like exoplanet can also characterize the atmospheres of hot desert worlds.
\end{itemize}

The first confirmed habitable exoplanet could be a hot desert world -- if such planets are frequent -- because it is easier to characterize their atmospheres. The reason is that close-in exoplanets orbiting small stars have a large transit probability, and a large transit over orbital period ratio that increases the observational window for a given mission. Characterizing low-mass desert worlds around nearby M dwarfs is especially favorable because the atmospheric scale height increases for low-mass planets while the planet-to-star area ratio remains large. Therefore, future modeling studies will aim to better understand the atmospheric circulation and the water cycle of such exoplanets. Observational campaigns conducted to characterize Earth-like exoplanet atmospheres should not neglect potentially rocky exoplanets closer to their host star than the inner edge limit suggests, and include hot desert worlds as potential candidates for life.

\acknowledgements
A. Zsom is grateful for discussions with Courtney Dressing, Vincent Eymet, Nikole Lewis, Renyu Hu, Ramses Ramirez, Daniel James Cziczo, Ray Pierrehumbert, James Kasting, Rory Barnes, and Felipe Gerhard. A. Zsom also thanks Ian Crossfield and Edwin Kite for providing detailed feedback on the first version of the manuscript. We thank the anonymous referee who helped to significantly improve the quality of the paper. A. Zsom was supported by the German Science Foundation (DFG) under grant ZS 107/2-1. J.d.W. acknowledges support from the B.A.E.F. (Belgian American Educational Foundation) and the W.B.I. (Wallonie-Bruxelles International) in the form of fellowships. V.S. was supported by the Swiss Science Foundation (SNF) under grant PBSKP2\_143496.

\appendix

\section{Model validation}
\label{app:modval}
\paragraph{Thermal emission}
We compare our radiative transfer code against the radiative transfer (RT) module of the Community Climate Model (\emph{ccm}3) of the National Center for Atmospheric Research \citep{Kiehl1998}. The \emph{ccm}3 model was developed for climate change studies, thus its radiative transfer module is more accurate than our code in the temperature, pressure, water and CO$_2$ mixing ratio ranges typically encountered in the atmosphere of Earth. Several effects are included in the \emph{ccm}3 RT model that we ignored to keep our code simple. Some of these effects are: properly treating overlapping lines of different absorbers (we use the assumption of random overlapping), calculating the IR flux in subdivided atmospheric layers to resolve the temperature gradient through the layer (we assume that the layers are isothermal). We show by the end of this section that ignoring these effects results in a relative error on the order of a few percent in the thermal IR fluxes. 

\begin{figure}
\centering
\includegraphics[width=0.49\textwidth]{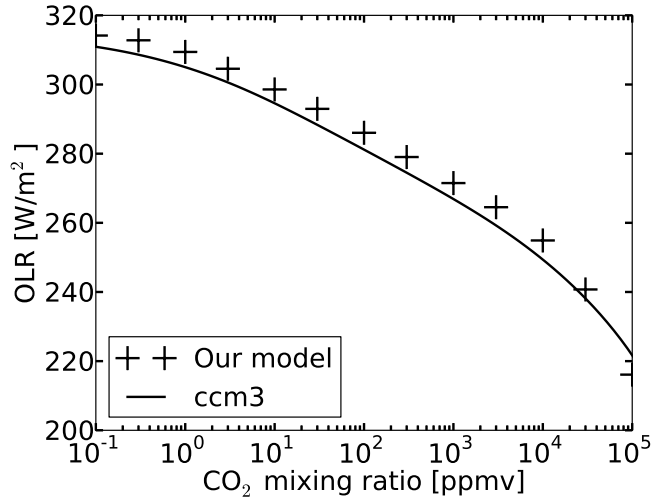}
\caption{The outgoing thermal flux of a dry N$_2$-O$_2$ atmosphere with 1 bar surface pressure, 273 K surface temperature, and various CO$_2$ mixing ratios. The solid line represents calculations performed with the Community Climate Model (\emph{ccm}3) of the National Center for Atmospheric Research. Thermal flux values calculated with our model are shown with `+' signs. The relative error between our results and the \emph{ccm}3 model is below 2\%. \label{fig:CO2}}
\end{figure}

\begin{figure*}
\centering
\includegraphics[width=0.49\textwidth]{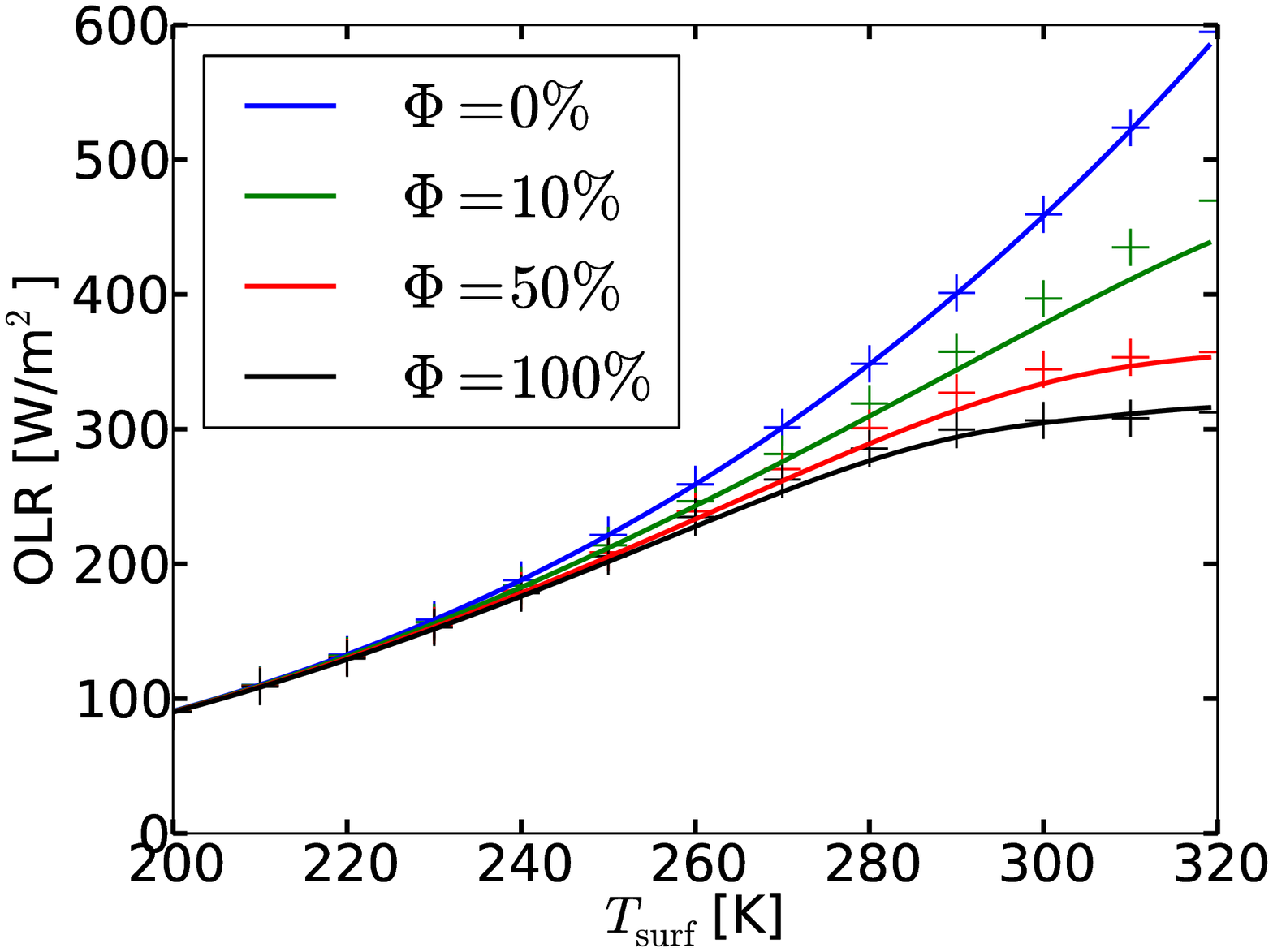}
\includegraphics[width=0.49\textwidth]{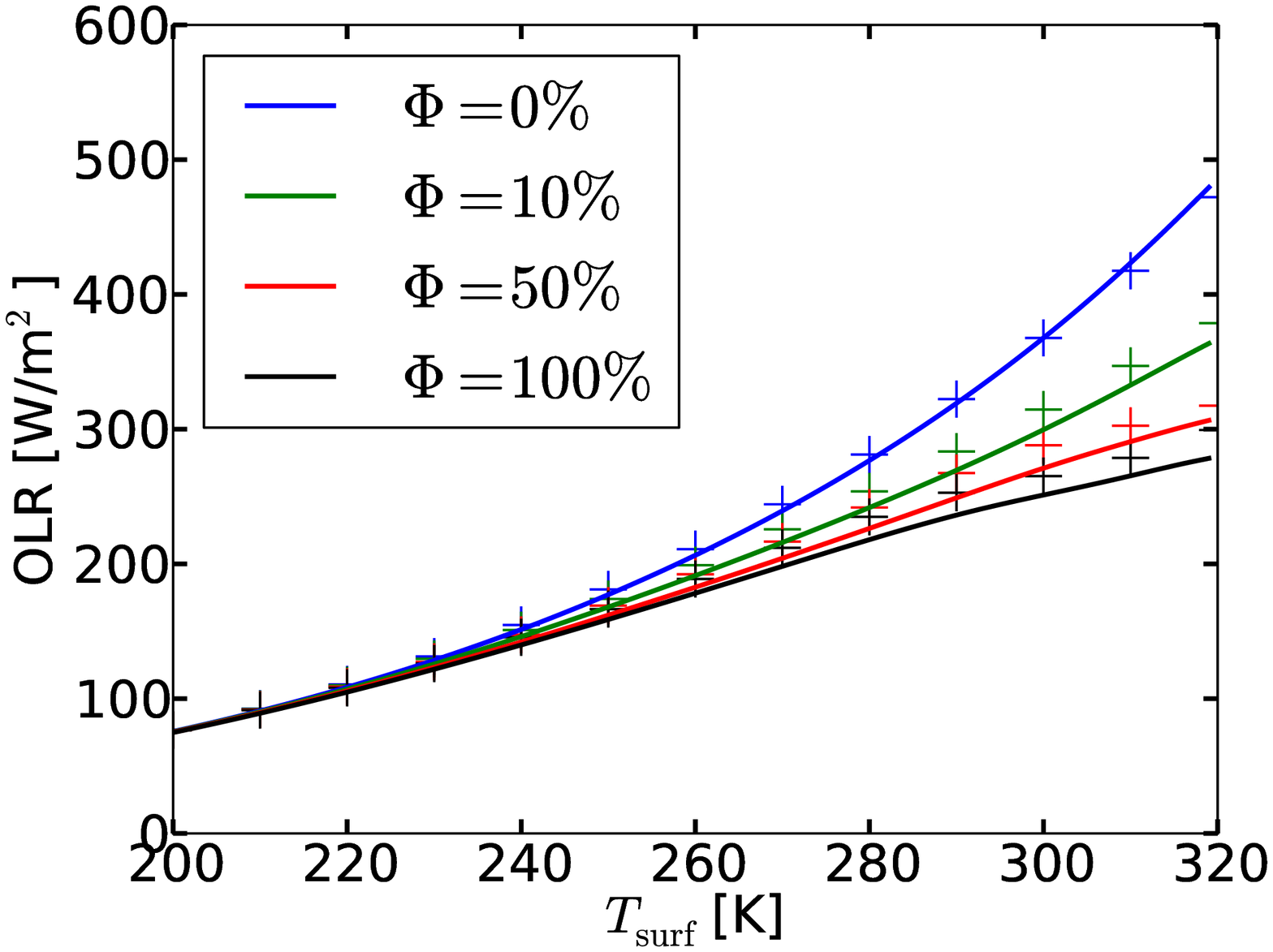}
\caption{\emph{Left:} The outgoing thermal flux of a moist adiabat atmosphere for four levels of relative humidity and various surface temperatures without CO$_2$. \emph{Right:} The outgoing thermal flux of moist adiabat atmosphere profiles with 10$^4$ ppmv CO$_2$ added. The solid lines correspond to \emph{ccm}3 calculations, the `+' symbols show our results. The relative error is less than 7\% in all cases. \label{fig:H2O}}
\end{figure*}

We calculate the outgoing thermal emission ($OTE$) of all-troposphere atmospheres with various temperatures, relative humidities and CO$_2$ mixing ratios using our model and \emph{ccm}3 RT. The first comparison is performed using a dry N$_2$-O$_2$ atmosphere with 273 K surface temperature and various CO$_2$ mixing ratios. The results are shown in Fig. \ref{fig:CO2}. The solid line represents the \emph{ccm}3 RT results, the `+' signs correspond to our calculations. Although our fluxes are typically larger than the \emph{ccm}3 fluxes, the relative error between our results and \emph{ccm}3 is less than 2$\%$, thus our model agrees well with \emph{ccm}3.

Next, the $OTE$ for an N$_2$-O$_2$ all-troposphere atmosphere is calculated with four different values of relative humidity: $\Phi$ = 0, 10, 50, and 100 $\%$, but without CO$_2$ (see Fig. \ref{fig:H2O}a). If the relative humidity is zero, there is no greenhouse gas in the atmosphere, thus the corresponding fluxes are generated by the black body emission of the surface. At surface temperature below 240 K, the water content of the atmosphere is so low that only a small amount of the IR flux is held back. Fig. \ref{fig:H2O}a shows how increasing the relative humidity lowers the outgoing thermal emission at the top of the atmosphere. The maximum relative error observed between our and the \emph{ccm}3 fluxes is 7$\%$, which we find acceptable.

The next level of complexity is to include both water vapor and CO$_2$ in the atmosphere. Fig. \ref{fig:H2O}b shows the $OTE$ of the atmosphere profiles used in the previous paragraph with 1\% CO$_2$ added. The outgoing thermal flux is significantly reduced by CO$_2$. As before, the maximum relative error is an acceptable 7$\%$.


\end{document}